\documentclass[preprint,12pt]{elsarticle}

\usepackage{graphicx}
\usepackage{epsfig}

\usepackage{amsmath}

\journal{Astroparticle Physics}

\begin{document}
\newcommand{\ecos}{E_\mu^{surface} \cos \theta}
\newcommand{\emu}{E_\mu^{surface}}
\newcommand{\kpm}{K^+/K^-}
\newcommand{\pipm}{\pi^+/\pi^-}
\newcommand{\ec}{E_\mu^{surface} \cos \theta}
\newcommand{\ecc}{E_\mu \cos \theta}

\begin{frontmatter}




\title{Interpretation of the Underground Muon Charge Ratio
}

\author[anl,ben]{P. A. Schreiner}
\author[anl] {J. Reichenbacher \fnref{ua}}
\author[anl] {M. C. Goodman}
\address[anl]{High Energy Physics Division,
Argonne National Laboratory, 9700 S. Cass Ave. Argonne, IL 60439, USA}
\address[ben]{Department of Physics, Benedictine University,
5700 College Road, Lisle, IL 60532, USA}

\fntext[ua]{Now at: The Department of
Physics and Astronomy, Box 870324, University of
Alabama, Tuscaloosa, Alabama 35487 USA}

\begin{abstract}

The MINOS experiment has observed a rise in the underground
muon charge ratio $r_\mu$ = ${\mu^+/\mu^-}$.  This ratio can be related to
the atmospheric production ratios of ${\pi^+/\pi^-}$ and ${K^+/K^-}$.
Our analysis indicates that the relevant variable for studying the charge ratio is 
$\ecos$, rather than $\emu$.  We compare a simple energy dependent
parameterization of the rise in the charge ratio with 
more detailed previously published Monte Carlo simulations and
an analytical calculation.
We also estimate the size of two previously neglected effects in
this context: the charge sign dependency of the dE/dx in rock, and
the energy dependence of heavy primaries on the derived ${K^+/K^-}$ ratio.

\end{abstract}

\begin{keyword}
Underground Cosmic Ray Muons \sep Muon Charge Ratio \sep Meson Charge Ratio
\PACS 13.85.Tp \sep 13.85.Ni

\end{keyword}

\end{frontmatter}


\section{Importance of Charge Ratio Measurements}
Atmospheric muons come dominantly from the decay of $\pi$s and $K$s 
produced in hadronic showers when
cosmic rays interact in the earth's atmosphere.  These muons have
been studied with energies ranging from hundreds of MeV to well
over a TeV.  A quantitative understanding of cosmic ray muons 
has value for a number of diverse topics, from atmospheric neutrinos
to the chemical composition of the highest energy cosmic rays.
The charge ratio of cosmic ray muons has been previously measured
over three orders of magnitude in energy.  Recently the MINOS
experiment\cite{bib:prd,bib:nd} presented data that for the first time
showed a rise in the measured charge ratio 
\begin{equation}
r_\mu \equiv \frac{N^{\mu+}}{N^{\mu-}}
\end{equation}
from previous measurements at high
values of $E_\mu$ or more specifically, high values of $\ec$.
\par In this paper, we discuss some of the issues involved in the
measurement and interpretation of the muon charge ratio.
In particular, we develop a simplified model where the rise
in the charge ratio can be understood from the properties of
$\pi$ and $K$ mesons, and the observation of the rise can be
used to determine the $\pipm$ and $\kpm$ ratios.
We also address several other issues related to the measurement 
of the charge ratio,
including the role of muon energy loss, a detector's Maximum
Detectable Momentum (MDM), and the effect of possible differences
in the spectral index of cosmic ray Hydrogen and Helium on the
interpretation.

\par Since the primary cosmic rays are
mostly positively charged protons, more secondary $\pi^+$ are expected than
$\pi^-$.  The quark content of the protons and of the atmosphere has been
used to estimate the $\pi^+/\pi^-$ ratio to be near 1.27 \cite{bib:naumov}.    
The charge ratio for kaons is even higher due to the phenomenon of
associated production.  
Strange particle production starts with the creation of an s quark
and an $\bar{s}$ quark.  An s quark which ends up in a nucleus is
associated with an $\bar{s}$ quark in a $K^+(\bar{s}u)$.  The $\bar{s}$
quark will not be in a baryon.  There is also $K^+K^-$ pair production.
Phase space favors 
hadronic production of $K^+\Lambda$ over $K^+K^-$ pairs at all
energies, so large $K^+/K^-$ ratios are expected.

\par  A standard parametrization of the atmospheric muon 
energy spectrum is given by Gaisser\cite{bib:gaisser}:
\begin{equation}
\frac{dN_\mu}{dE_\mu} = \frac{0.14 E_\mu^{-2.7}}{\rm cm^2~s~sr~GeV}
\times 
\left \{ 
\frac{1}
{1+\frac{1.1 E_{\mu}\cos \theta}{\epsilon_\pi}} 
+\frac{\eta}    {1+\frac{1.1 E_{\mu}\cos \theta}{\epsilon_K}}
\right \} 
\label{eq:gaisser}
\end{equation}
The two terms inside the curly bracket give the contributions from charged pions and 
kaons.
For the
hadron mass ($m_i$) and lifetime ($\tau_i$), and the atmospheric
scale height (h $\sim$ 6.4 km), the values of
 $\epsilon_i = m_i ch/\tau_i $ are the energies where the 
probability of meson interaction in the atmosphere and decay are equal:
$\epsilon_\pi$=115 GeV and $\epsilon_K$= 850 GeV\cite{bib:zatsepin}.
The zenith angle is denoted by $\theta$. In those experiments with a flat overburden, the largest muon
intensity is at $cos({\theta})=1$.  Reference \cite{bib:gaisser}
gives $\eta$ = 0.054, which sets the relative
contribution to the muon flux from $\pi$ and K decay.
The parameter $\eta$ depends upon the 
$\pi/K$ ratio, branching ratios, and kinematic factors which arise due
to differences between the $\pi$ and K masses. 
 \par
The ratio between the
two terms in Equation~\ref{eq:gaisser} quantifies the relative contribution
to the muon flux from muons due to ${\pi}$ and K decay.  It depends 
on $\ec$ and rises from $\eta$ = 5.4\% at low energies to 
$\eta \times \epsilon_K/\epsilon_{\pi}$ = 40\% at high $\ec$. This substantial 
rise which occurs mostly between 115 GeV and 850 GeV
is caused by the different masses and lifetimes of pions and kaons,
not by any increased amount of kaon production from
primary cosmic ray interactions in the atmosphere. 
\par 
MINOS has provided the first high statistics measurements of
the muon charge ratio at $\ec > 115$ GeV.  We will show that this gives
the needed sensitivity for extracting information about cosmic
ray produced pions and kaons
separately.  We also note that
a similar calculation for neutrinos shows that kaons are the dominant
parent for TeV neutrinos, which
are the largest backgrounds for astrophysical source searches
at neutrino telescopes.

\par The organization of this paper is as follows:  In the following
section, we review several measurements of the atmospheric muon
charge ratio.  In Section ~\ref{sec:mdm}, we discuss some of the
issues involved in measuring the charge ratio in a magnetic detector
underground.  In Section~\ref{sec:eloss}, we address particular issues related to the 
muon energy loss and how the muon energy at an underground detector is
related to the energy in the atmosphere.  In Section~\ref{sec:models},
we use Equation~\ref{eq:gaisser}, which was based on ideas developed
by Zatsepin\cite{bib:zatsepin}, to parameterize the charge 
ratio in a simplified model.  Zatsepin's work is
the earliest reference that we have found which defines the roles
of the critical energies $\epsilon_\pi$ = 115 GeV and $\epsilon_K$
= 850 GeV.  There are two important implications from this model.  First, the
charge ratio depends not simply on the muon energy $\emu$, but on the
combination of the energy and zenith angle $\ec$.  Second, 
together with previous measurements at low $\ec$, 
observation of the rise in the charge ratio, 
which mainly occurs between $\epsilon_\pi$ and 
$\epsilon_K$, can be used to fit the meson production charge
ratios $\kpm$ and $\pipm$.  
In subsequent sections we discuss how this model compares to several
full Monte Carlo calculations of the muon charge ratio, and how
a possible difference in the spectral index of the primary cosmic Hydrogen and Helium
flux might affect the interpretations.

\section{Compilation of Measurements of the Atmospheric Muon Charge Ratio}
\label{sec:previous}
 Numerous measurements of the atmospheric muon charge ratio data
 have been made
 at the earth's surface, from MeV to multi-TeV energies. The Hebbeker 
 and Timmermans 2002 compilation 
 article \cite{bib:hebbker} provides many  references to these data.
 Figure 1 displays muon ratio data as a function of the surface energy;
 data from six experiments in the energy range
 0.10 TeV to 10 TeV are shown. 
 \par
Baxendale et al. \cite{bib:baxendale} published extensive muon flux 
data 
in the momentum range 7-500 GeV/c using
the Durham spectrometer.  This detector consisted
 of magnetized iron blocks, flash
tubes, and scintillation counters. The magnetic field in the three 
magnetized blocks was reversed regularly to reduce bias effects.
Their highest energy charge ratio data point is at 358 GeV/c. 
The publication did not give the ratio as a function of zenith angle, and
thus only the energy dependency of the charge ratio is used here. 

 \par
The CosmoALEPH experiment \cite{bib:cosmo} at the LEP facility at CERN
published 8 data points in the 80-1600 GeV surface energy range. The
cosmic ray portion of the experiment used only the hadron calorimeter
and the time projection chambers. The detector was located at 
a depth of 320 m.w.e. Apparently the magnetic field was
not reversed during data collection. The zenith range of 0-10 $\deg$
 was studied, but the
publication did not give the ratio as a function of zenith angle.
Thus only the energy dependency of the charge ratio is used here. 

\par
The L3+C experiment\cite{bib:l3} has published extensive muon charge ratio
data at many zenith angles at momentum up to 380 GeV/c.  The high accuracy
and extensive angular range of this data are ideal for comparison with
our parameterization of the muon charge ratio.  We have used 
this data extensively in Section 4 of this paper.

\par
Matsuno et al. \cite{bib:matsuno} published data in the momentum range from 70 GeV/c to 20 TeV/c 
using the MUTRON magnetic cosmic ray spectrometer.  It consists of
a solid iron magnetic spectrometer, a calorimeter of 16 proportional
chambers, and 48 spark chambers. Data were collected with both
forward and reversed magnetic fields.  The data were collected only at large
zenith angles 86-89 degrees.  While the data sample is large, there
are only a few hundred events above 2 TeV/c. 
 \par
 MINOS has published data \cite{bib:prd,bib:nd} from both their 
Far Detector at Soudan, Minnesota and their Near Detector at Fermilab. Data from 
both detectors were collected with
 both forward and reversed magnetic fields. Data are available as a function
 of the zenith angle.
 \par
Rastin \cite{bib:rastin} published data in the muon surface
energy range from 6 - 1288 GeV
using the Nottingham solid iron magnetic spectrometer. Data were 
collected with both forward and reversed magnetic fields.

\par
Experiments with data points with relatively large error bars have not
been included in this analysis since they are not useful for the 
precise studies in this paper.  Experiments with charge ratio data
values at energies above 100 GeV, with all data points having quoted errors 
greater than +-0.1, have not been included
in our analysis; thus Hayman and Wolfendale \cite{bib:hayman}, 
Appleton et al.\cite{bib:appleton}, 
Nandi et al.\cite{bib:nandi}, 
and Kremer et al. \cite{bib:kremer} are not included.
Specifically, Burnett et al. \cite{bib:burnett} was not included in our
 study due to the absence of any
published estimates of systematic error on their ratio values.

\par
\begin{figure}
\begin{center}
 \includegraphics [width=5.0in]{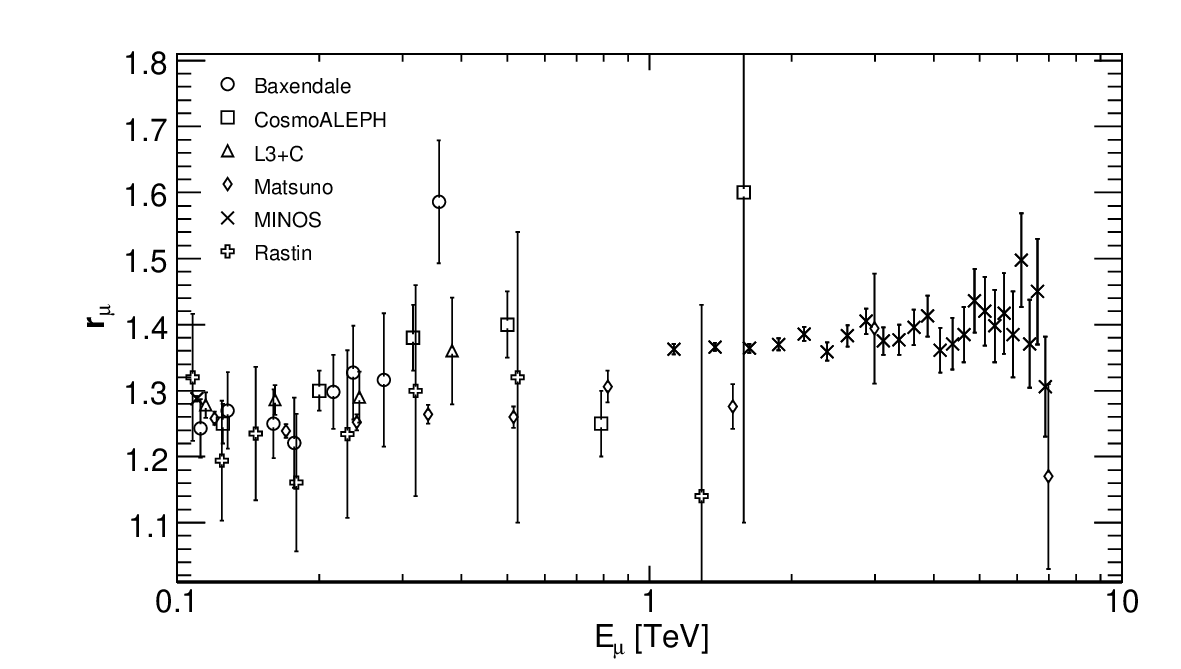}
 \caption{\label{fig:otherexps} Compilation of muon 
charge ratio from 
 experiments with at least one 
data point with an uncertainty less than 0.10.
}
 \end{center}
 \end{figure}

 As seen in Figure~\ref{fig:otherexps}, the ratio has an
energy dependency; it is near 1.25 near 100 GeV
 and rises to a value near 1.4 at several TeV.
  The rise in the ratio has been attributed to
 an increasing  contribution from kaon decay\cite{bib:naumov}. 
 The data in Figure \ref{fig:otherexps} could be
 parameterized with a linear function of $\emu$ or a log($\emu$) function,
but there is no physics reason to expect a linear dependency with E, or
 even a log(E) dependency.  We have chosen to use a different parameterization that
 encompasses more of the kinematics and the physics of the muon 
 production processes, as will be described in Section~\ref{sec:models}.

\section{Issues in Measuring the Charge Ratio}
\label{sec:mdm}
\par In this section, we review the method used for determining
the muon charge, and consider some of the
systematic errors which affect such a measurement.  We describe
the concept of Maximum Detectable Momentum, and consider two
distinct types of systematic errors:  bias and randomization.

\par Most measurements of the atmospheric muon charge ratio involve
the use of a magnet to deflect the trajectory of a charged particle.
The resulting curvature is used to measure both the charge sign and
momentum of the muon.  Depending on the size, granularity, and
strength of the magnetic field, every detector has a Maximum
Detectable Momentum (MDM).  We define the MDM as that momentum
for which a nearly straight real track will have a measured
curvature (determined from a fit to points along the track) 
which is one standard deviation from zero.  For a uniform magnetic field, the MDM is
simply the reciprocal of the error (s.d.) of the curvature
measurement, when the curvature is expressed in (GeV/c)$^{-1}$.
\par The MDM can be calculated for each event in a
detector.  As a simplified but instructive example, we consider a simplified
toroid detector which is a long right circular cylinder of radius R,
with uniform density, with a B field that is
azimuthal and constant.  The long z-axis is perpendicular to the
zenith direction.  We do not consider any
tracks which enter the front or leave the rear
of the cylinder, i.e. all muons enter and exit the side.
  All such $\mu$ tracks will be roughly S shaped, with curvature
one way for the first half of the track, and the other way for
the second half. 
The length of the corresponding half-track is
\begin{equation}
L = \sqrt{R^2-b^2}/\sin(\psi)
\end{equation}
where b is the impact parameter or distance of closest approach
to the center of the toroid, and $\psi$ is the
angle of the track with the z axis (not the zenith angle). For this
discussion, $\psi$ is near 90 degrees.
To calculate the precision of the momentum measurement, we need the
component of the magnetic field B which is perpendicular to the
track direction.  Approximating at the midpoint of the half-track,
assuming that $\psi$ is not small, we find
\begin{equation}
B_{PERP} = B \times \sqrt{[1-\sin^2(\psi)\times \cos^2
(\arctan(\sqrt{(R^2-b^2)}/2b))]}
\label{eq:bperp}
\end{equation} 
Equation~\ref{eq:bperp} is used in Reference \cite{bib:matsuno}, and in Reference \cite{bib:mdm} to calculate the
MDM for MINOS, which varies dramatically between 470 GeV 
to as low as 15 GeV for large impact
parameters and large $\psi$, where many muon tracks occur.
\par 
The above discussion deals with the ability to measure charge
in an ideal detector.  In a real detector, there are a number of
reasons that we might assign an incorrect charge sign to a track.  We
separate these errors into two kinds which we call bias and randomization.  
One kind of error involves biases towards one sign
or the other.  Other errors will cause tracks to be randomly
assigned as positive or negative.  These randomized events will have
an apparent charge ratio of unity.  Since the true charge ratio is
greater than one, any such errors will lower the measured charge 
ratio.   
We next describe two instances of each kind of error.

\par The first kind of bias is associated with the magnetic
geometry of
the detector, which we call acceptance effects.   
Ignoring the very small magnetic field in the air outside the detector,
the charge ratio of tracks which enter the detector is not affected by the
magnetic field and/or the geometry of the detector.  Instead, there are 
effects which have to
do with cuts that are made to require a good track, and the
efficiency of those cuts.  From a given direction, tracks of one
kind (the other) will be bent such that the track length in the
detector is longer (shorter).  In order to require a minimum number
of hits to make a good charge determination, a minimum length cut 
may be made.  Thus tracks in some directions may be biased towards
one sign or the other.  
In the MINOS geometry, such directions 
occur as peaks or dips in the azimuthal distributions of the charge
ratio.  They tend to be symmetric with respect to the detector, but
since a varying overburden can cause more tracks from some directions
than others, they do not necessarily cancel.  
\par A second kind of bias comes from any misalignment of detector 
planes which causes a curvature in the coordinate system.  A result
of any such misalignment would be that straight tracks appear to have
curvature.  With the MINOS geometry, it was found that a systematic
misalignment of even a fraction of a mm (as measured by the track's
sagitta) would have a noticeable effect on the charge ratio.  
A curvature cut does not remove such a bias.
\par Bias effects as described above are correctable in principle,
with an accurate Monte Carlo.   Acceptance issues can be corrected
if the model of the magnetic field is accurate, and if the overburden
is correctly modeled so that the expected fluxes as a function of
zenith and azimuthal angle are known.  Misalignments can be modeled
in principle, but must be done with a correct three-dimensional model
of the detector, with an accuracy that
is extremely difficult to achieve in practice. 
\par
An easier way to deal with the above bias effects is to reverse the magnetic
field.  If the reversal of the field is done with sufficient 
accuracy, then all forms of bias described above cancel when
one uses the geometric mean of the charge ratio for
forward and reversed field running \cite{bib:matsuno}
\begin{equation}
r = [r_{forward} \times r_{reversed}]^{1/2}
\label{eq:gmean}
\end{equation}
This cancellation is apparent when one considers the ratio of the
positive muons in forward(reverse) running to 
negative muons in reverse(forward) running.
The equality of these two values is a powerful
consistency check.
\par By contrast, effects which tend to randomize the charge ratio do not
typically cancel in this manner and must be dealt with separately.
An obvious type of randomization occurs for high momentum straight
tracks above the MDM.  
If a track is straight within the resolution 
of the detector, then we will be unable to accurately determine the
charge sign.  For a straight enough track, a program will measure 
curvature either way an equal fraction of the time.   The fit
that calculates the curvature can also calculate the error on
the curvature.  This form of randomization can be minimized
by cutting on the fractional curvature error $\sigma(1/p)/(1/p)$ as 
is shown in
Figure 5 of Reference~\cite{bib:prd}.
However the fraction of mismeasured tracks may not follow a
Gaussian distribution for $\sigma(1/p)/(1/p)$ and Monte Carlo
estimates of the misidentification must be analyzed with care.
\par A second form of randomization, which was observed in MINOS data,
was generally associated with extra hits.  These might be due to
noise, crosstalk or demultiplexing issues, or also delta rays or
other particles associated with the muon.  For beam experiments,
the magnetic field is perpendicular to the beam direction, to 
maximize bending power from the magnet.  For cosmic
rays, muons are observed at all angles.  As described above, muon tracks
will be bent much more in a toroidal magnet if they impinge on
the detector with a low impact parameter.  For those cosmic ray muons
which are traveling mostly parallel to the magnetic field (at large
impact parameter), an
interesting situation can arise if the apparent bending due to multiple
scattering is comparable to the bending from the magnetic field.
If that happens, a reconstruction program which takes into account
multiple scattering can get a good fit for a straight track for
any momentum.  The inclusion of a hit or hits which do not belong on 
the track can then lead to a class of muon tracks which are really
at high momentum, but get reconstructed at low momentum with a random
charge sign depending on which side of the track the extra hits appeared.
\par We can quantify this result as follows:  Projected onto a plane,
the RMS multiple Coulomb scattering angle can be written
\begin{equation}
<\theta_{MCS}> = [0.0136/p(GeV/c)]*\sqrt{(L(meter)/X)}
\end{equation}
where X is the radiation length.
The magnetic bending for this same length of track is given by
\begin{equation}
\theta_{Bend} = L(meter)*B_{PERP}(Tesla)/[3.336 \times p(GeV/c)]
\end{equation}
Note that in the ratio $\theta_{MCS}/\theta_{Bend}$, p cancels. For typical muon tracks
in MINOS, the ratio rises from 0.14 at
low impact parameter and $\psi$  to near unity at large impact
parameter.  Thus the type of randomization
described in the previous paragraph can be
preferentially removed by discarding events with large values
of $<\theta_{MCS}>/\theta_{Bend}$.

\par The cancellation of bias effects can be checked by
studying distributions which have bias effects in them before
and after the forward and reverse field data are combined.  
We have no such consistency check for randomization effects.
Thus, the
ability to detect and control all bias effects using the combined forward
and reverse field data is stronger than the ability to cancel all
randomization effects.  This might lead to a tendency for experiments to 
report a charge ratio which is systematically smaller than the
true value.  We also note that not every experiment listed in
Section~\ref{sec:previous} has taken data with a reversed magnetic
field.

\section{Issues with Interpreting Data Obtained Underground}
\label{sec:eloss}
\subsection{The Relevance of Energy Loss Differences}
\par The charge ratio measured in underground experiments
is a ratio
of $\mu^+$ to $\mu^-$ at the detector.  For comparison with production
models, the relevant quantity is the ratio in the upper
atmosphere.  If the energy loss of positive and negative muons were
identical, as they go through the earth, these two ratios would be
the same.  The leading energy loss processes of $\mu^+$ and $\mu^-$ are the
same of course, but there are corrections of the order of the fine
structure constant $\alpha$.
We evaluate these small differences, 
first for ionization 
energy loss, and then for radiative processes.
The latter difference
is found to be negligible.  The small difference in ionization energy
loss is used to calculate the difference in range and the 
correction to the charge asymmetry as a function of
slant depth in Section \ref{sec:asym}.
\par
The statistical energy loss of muons, traversing an amount $X$ of matter in 
$g/cm^2$, with energies far above the Bethe-Bloch minimum is usually 
parameterized as
\begin{equation}
- \frac{dE_{\mu}}{dX} = a(E_{\mu}) \,+\, 
\displaystyle\sum_{n=1}^{3} b_{n}(E_{\mu}) \cdot E_{\mu},
\label{eq:eloss}
\end{equation}
where $a$ is the collisional term (i.e. ionization) 
and $b$ in the second term accounts for the three radiative muon 
energy loss processes: \,Bremsstrahlung, \,pair production
and \,photo-nuclear interactions. In  
Table \ref{tab:1}\cite{bib:Groom}\cite{bib:pdg} these energy loss 
parameters are listed for standard rock. The critical energy 
where ionization 
losses equal radiative losses in standard rock is approximately $0.6\,$TeV. 
The average muon surface energy for a muon which reaches 2000 mwe is 
greater than $1$\,TeV, so the $b$ term and its energy dependence are important 
in calculating the energy loss. We have investigated differences 
in the $a$ and $b$ terms for $\mu^+$ and $\mu^-$. 

\begin{table}[ht]
\begin{center}
{
\begin{tabular}{|c||c||c|c|c|c|}
\hline
$E_{\mu}$ & $a_{ion}$ & $b_{brems}$ & $b_{pair}$ & $b_{DIS}$ & ${\Sigma} b$ \\
\cline{3-6}
 [$GeV$] & [$MeV\,cm^{2}/g$] &\multicolumn{4}{c|}{$10^{-6}\,cm^{2}/g$} \\
\hline\hline
$10$ & 2.17 & 0.70 & 0.70 & 0.50 & 1.90 \\
$10^{2}$ & 2.44 & 1.10 & 1.53 & 0.41 & 3.04 \\
$10^{3}$ & 2.68 & 1.44 & 2.07 & 0.41 & 3.92 \\
$10^{4}$ & 2.93 & 1.62 & 2.27 & 0.46 & 4.35 \\
\hline

\end{tabular}
}
\caption{{\small Average muon energy loss parameters calculated for 
standard rock \cite{bib:Groom}\cite{bib:pdg}}}\label{tab:1}
\end{center}
\end{table}

\subsection{Difference in Ionization dE/dx for $\mu^+$ and $\mu^-$}

\par
\begin{figure}
\begin{center}
\noindent
\includegraphics [width=0.5\textwidth]{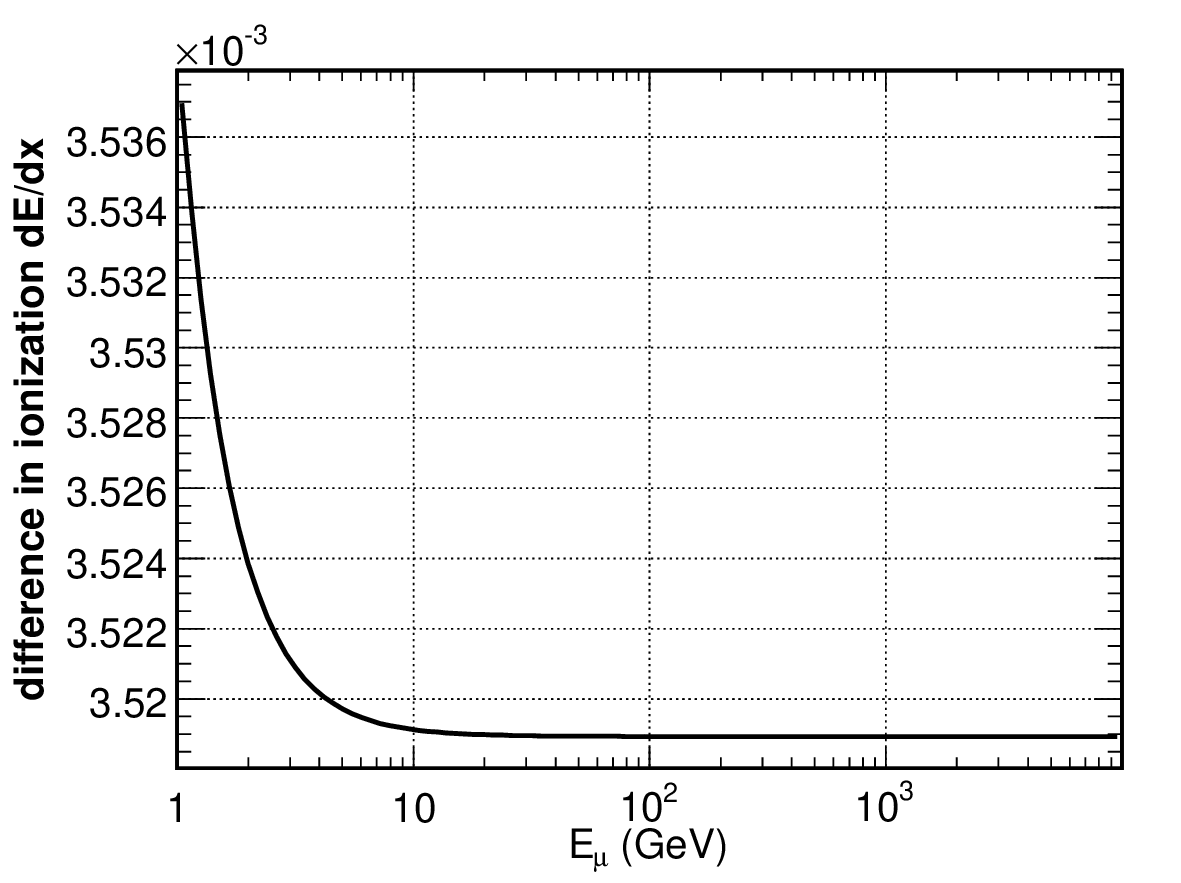}

\end{center}
\caption{Calculated difference in ionization energy loss between
positive and negative muons in standard rock (average nuclear properties:
$\overline{Z}=11$, $\overline{A} = 22$ \,\cite{bib:pdg}).}
\label{fig:fig1}
\end{figure}

 At low energies, around the Bethe-Bloch minimum, the difference in 
ionization energy loss is known as the Barkas effect\cite{bib:fermi}, 
and there have been efforts to both measure and calculate 
the difference\cite{bib:mccarthy}. 
Calculations show that negative particles lose energy at a slower 
rate, with the difference dropping from tens of percent at MeV energies 
to about $0.3$\,\% in the GeV range. 
Such differences were experimentally verified both at MeV 
energies\cite{bib:barkas1}\cite{bib:barkas2}\cite{bib:barkas3} 
and in the GeV range \cite{bib:barkas4}. 
At higher energies, 
this difference in ionization energy loss has usually been neglected, 
and we are not aware of any measurements. As described in Reference 
\cite{bib:jackson}, the usual ionization 
energy loss term for muons (of either sign) 
depends on $z^2$, and the difference between $\mu^+$ and $\mu^-$ arises from 
a small additional $z^3$ correction term. 
This correction term in $dE/dx$ is:
\begin{equation} 
\displaystyle\left(\frac{dE}{dX}\right)^{corr}_{ion} = \frac{\pi \alpha z^3 0.307 Z}{2 \beta A} \quad 
[MeV\, cm^{2}\, g^{-1}]
\label{eq:3}
\end{equation}
where $\alpha$ is the fine structure constant, $z$ is the charge, $\beta$ is 
the relativistic velocity, and $Z$ and $A$ are the nuclear properties of the 
material through which the muon is passing. The absolute value of the 
difference in ionization energy loss between positive and negative muons in 
standard rock\cite{bib:pdg} is plotted in Figure \ref{fig:fig1}. 
It is fairly constant above $10$\,GeV, at a value corresponding to 
approximately $0.15$\,\% of the mean energy loss in the ionization dominated 
energy regime ({\em c.f.} Table \ref{tab:1})\cite{bib:icrc2}. 

\subsection{Calculated Difference in Bremsstrahlung dE/dX 
for $\mu^+$ and $\mu^-$}

\begin{figure}
\begin{center}
\noindent
\includegraphics [width=0.5\textwidth]{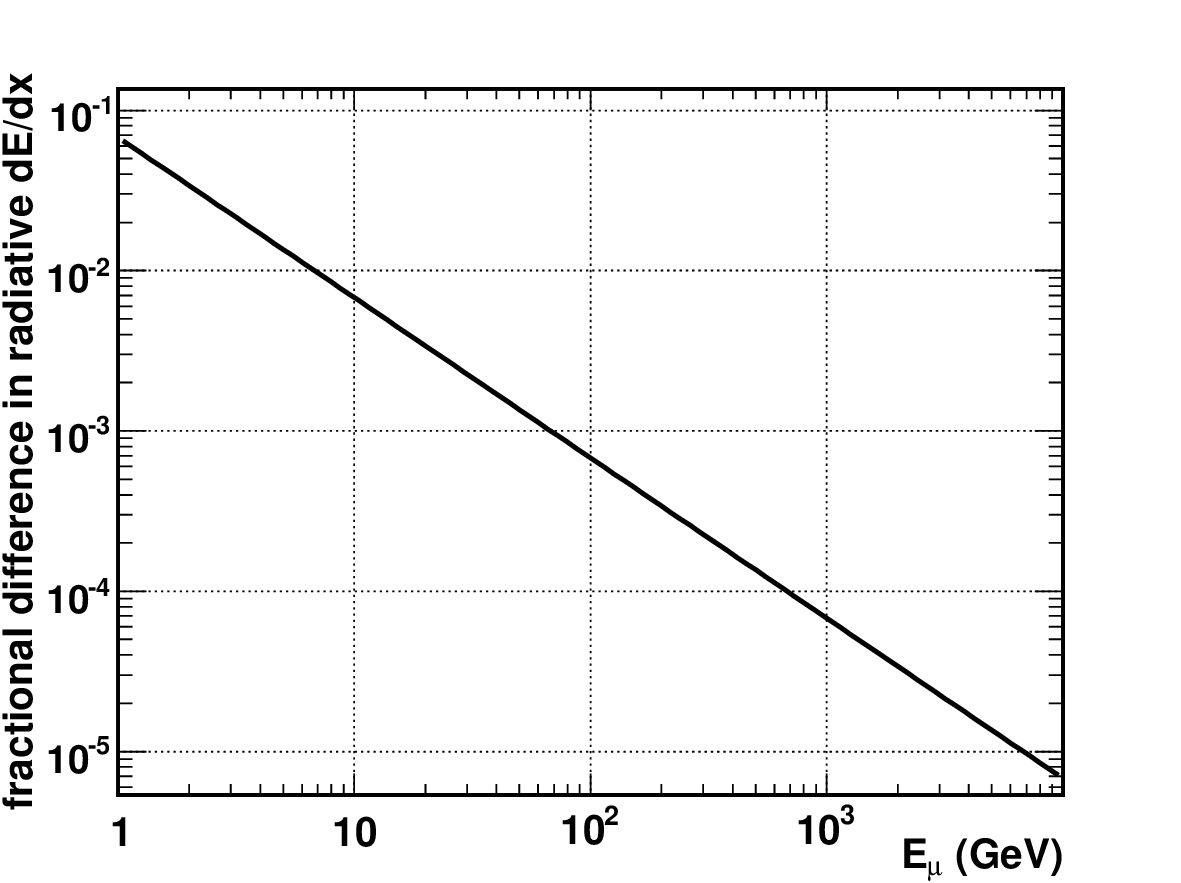}
\end{center}
\caption{Calculated fractional difference in Bremsstrahlung energy loss 
between positive and negative muons.}
\label{fig:fig2}
\end{figure}

Above an energy near $0.6$\,TeV in standard rock, radiative energy loss 
becomes comparable
to ionization energy loss, and continues to grow at
higher muon energies. 
Radiative energy loss has been calculated using the Eikonal
approximation in References \cite{bib:bk} and \cite{bib:lms}.
For the difference between positive and negative particles, the
leading term cancels, and only a term proportional to M/E = 1/$\gamma$
survives, which does depend on the sign of the charge\cite{bib:jackson2}:
\begin{equation}
\frac{[\frac{dE}{dX}]^{\mu+}_{brems} - [\frac{dE}{dX}]^{\mu-}_{brems}}{[\overline{\frac{dE}{dX}}]_{brems}} = \frac{8 Z \alpha}{\gamma}
\label{eq:4}
\end{equation}
where $\gamma$ is the Lorentz factor of the muon. 
Again, the $\mu^+$ has a slightly higher energy loss. 
This fractional difference {\it decreases} with energy and is already 
negligible where radiative energy losses become important. 
This fractional difference is plotted in Figure \ref{fig:fig2} for muons in 
standard rock. 
Presumably, the same fractional difference can also be assigned for 
pair-production, as the underlying process is a two-photon exchange between 
the muon and the constituents of the nucleus, and thus the cross sections for 
$\mu^+$ and $\mu^-$ should scale in the same way as for Bremsstrahlung. 

\subsection{Range Underground and Muon Charge Asymmetry}
\label{sec:asym}
\par Taking the vertical muon intensity from the Gaisser parameterization 
of the muon flux at the surface (see Equation \ref{eq:gaisser})
and propagating this energy spectrum 
underground according to statistical ionization and radiative energy losses, 
one can calculate the underground muon intensity. 
This procedure is described in detail in \cite{bib:myCrouchProc} for 
overburdens of standard and Soudan rock (MINOS). 
First, the average muon range underground, for each value of surface energy, 
is computed. For this, the energy dependent $a$ and 
$\Sigma b$ values were parameterized for standard and Soudan rock as in 
\cite{bib:myCrouchProc}. The additional ionization loss 
according to Equation \ref{eq:3} was added for $\mu^+$ to the value of the function for 
$a$ (subtracted for $\mu^-$). 
Radiative losses
($90\,\%$ of $\Sigma b$)
were scaled with the energy dependent fractional 
difference $8Z\alpha/\gamma$ according to Equation~\ref{eq:4} as $\Sigma b^{\pm} = 0.9 
\cdot \Sigma b \cdot (1 \pm 4Z\alpha/\gamma)$ for $\mu^+$ and $\mu^-$, respectively. 
The contribution from photo-nuclear production (DIS), 
which we are not aware is charge dependent, was not scaled.
For each initial value of muon energy,
the slant depth in meter-\-water-\-equivalent where the 
muons of different charge range out was determined.
\par
\begin{figure}
\begin{center}
\noindent
\includegraphics [width=0.5\textwidth]{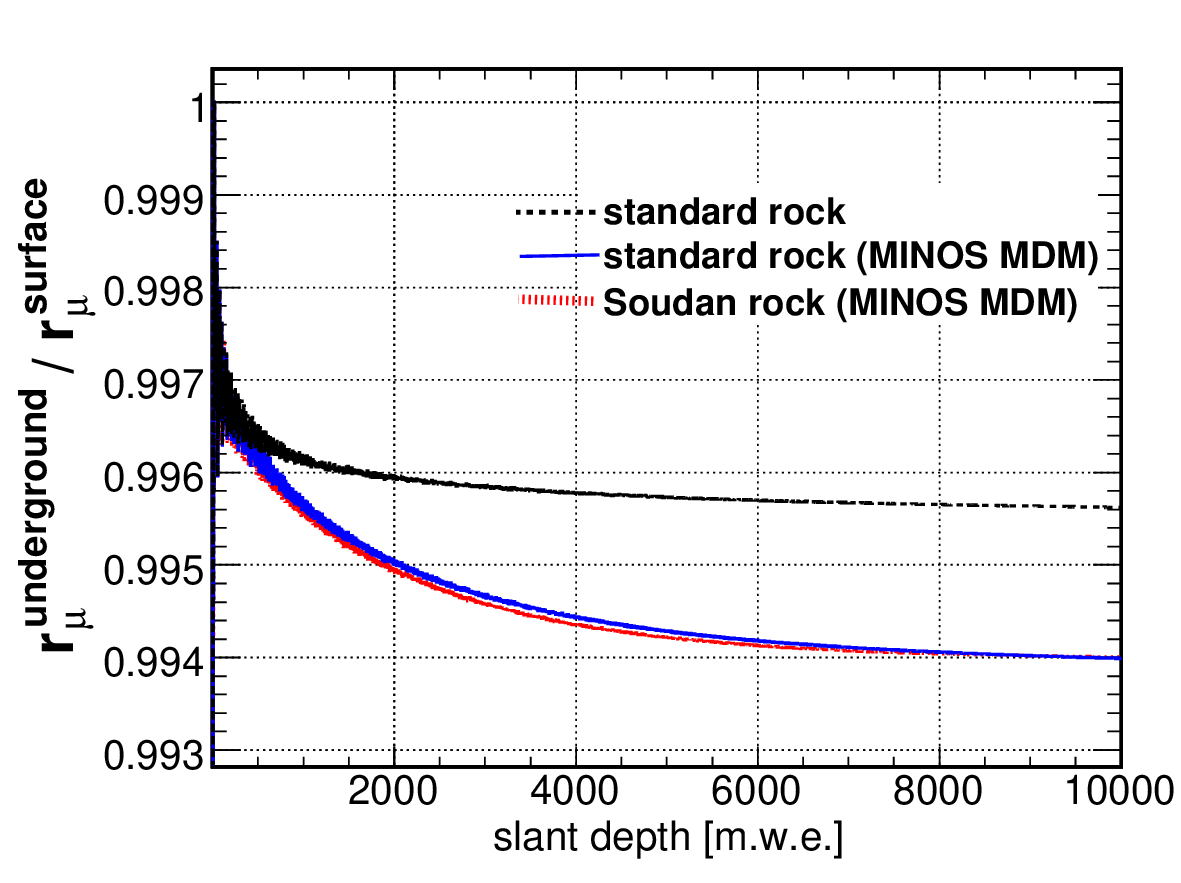}
\end{center}
\caption{\label{fig:calcratio}
Calculated ratio of positive to negative vertical muon intensities in 
Soudan rock as a function of slant depth for equal surface intensities. The upper curve is for all muons, 
the lower curve is for muons with a remnant momentum of less than 250 GeV/c 
($\approx$ the maximum detectable momentum in the MINOS Far
Detector).}\label{fig:intensity}
\end{figure}

 We computed the 
underground intensities of positive and negative muons as a function of slant 
depth for a given rock composition. 
This ratio of the $\mu^+$ and $\mu^-$ intensity curves is shown in 
Figure \ref{fig:intensity} for Soudan rock. The upper curve corresponds to the 
fractional difference in integral intensities of $\mu^+$ and $\mu^-$ at a 
given slant depth. For slant depth values above about $1000$ mwe the underground 
ratio $N(\mu^+)/N(\mu^-)$ is lowered by roughly $0.4\,\%$. 
The detected intensity 
corresponds to the charge ratio of the muons at depth below the MDM
for MINOS.  For 
increasing slant depth values the measured underground ratio 
$N(\mu^+)/N(\mu^-)$ is further  reduced and saturates at
about $0.6\,\%$ below its surface value for slant depths larger than 
roughly $5000$ mwe. 
\par 
This value can be qualitatively understood as follows.  If the muon
flux was proportional to $E^{-3.7}$, and if the muon range was strictly
proportional to energy, then less positive muons would survive underground
and
\begin{equation}
r_\mu^{underground} = r_\mu^{surface} 
\frac{(E_\mu - \Delta(E_\mu))^{-3.7}}{{E_\mu}^{-3.7}}
\end{equation}
So the change in the charge ratio is $\Delta(r_\mu) \sim 3.7 \Delta E_\mu/E_\mu$
Thus, the
small fractional difference in energy loss for $\mu^+$ and $\mu^-$ 
of the order of $0.15\,\%$ predicted by theoretical calculations at high 
energies, gets amplified by a factor of about $3.7$.
The impact of the rock composition is almost negligible, as the induced 
muon charge asymmetry under Soudan rock lowers 
the surface value of the ratio 
by an additional amount less than $0.02\,\%$ compared to standard rock. 
Figure \ref{fig:calcratio} can be used to
correct the underground measured muon charge ratio to its surface value.
For MINOS, this corresponds to a correction which increases
the measured value by 0.6\%.  

\section{Model of Pion and Kaon Contributions to the Charge Ratio}
\label{sec:models}

We have investigated a generalization of Gaisser's Equation~\ref{eq:gaisser} to 
 study separately
the positive and negative muon intensities.  
Using the positive fraction parameters $f_\pi$ and $f_K$, the energy
 dependency of the 
positive and negative muons is given by

\begin{equation}
\frac{dN_{\mu^+}}{dE_{\mu}} = \frac{0.14 E_\mu^{-2.7}}{\rm cm^2~s~sr~GeV}
\times 
\left \{ 
\frac{f_\pi}
{1+\frac{1.1 E_{\mu}\cos \theta}{115~\rm GeV}} 
+\frac{\eta\times f_K} {1+\frac{1.1 E_{\mu}\cos \theta}{850~\rm GeV}}
\right \} 
\label{eq:plus}
\end{equation}
\begin{equation}
\frac{dN_{\mu^-}}{dE_{\mu}} = \frac{0.14 E_\mu^{-2.7}}{\rm cm^2~s~sr~GeV}
\times 
\left \{ 
\frac{1-f_\pi}
{1+\frac{1.1 E_{\mu}\cos \theta}{115~\rm GeV}} 
+\frac{\eta\times(1- f_K)}  {1+\frac{1.1 E_{\mu}\cos \theta}{850~\rm GeV}}
\right \} 
\label{eq:minus}
\end{equation}
where $\epsilon_\pi$ and $\epsilon_K$ have
been replaced by their numerical values.
We can use equations \ref{eq:plus} and \ref{eq:minus} to calculate the
surface muon charge ratio:
{\large
\begin{equation}
r_{\mu} = \frac{\left \{ 
{\frac{f_\pi}{1 ~+~ {1.1 \ecc}/{115~\rm GeV} } 
~+~\frac{\eta\times f_K} {1~+~{1.1 \ecc}/{850~\rm GeV} }}
\right \} }
{\left \{\frac{1-f_\pi}{1~+~{1.1 \ecc}/{115~\rm GeV} } 
~+~\frac{\eta\times(1- f_K)}  
{1~+~{1.1 \ecc}/{850~\rm GeV}}\right \}}
\label{eq:pika}
\end{equation}
}

\begin{figure}
\begin{center}
\includegraphics [width=5.0in]{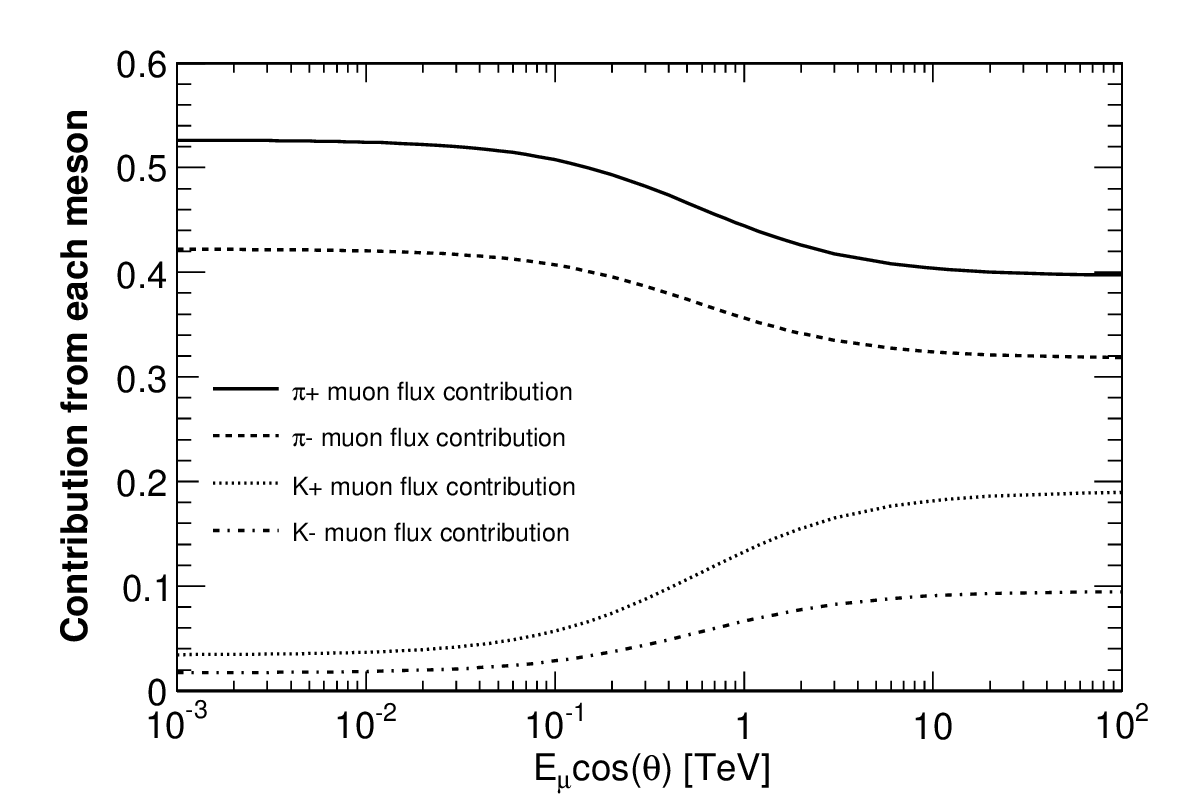}
\caption{\label{fig:mesoncontribution}
Contributions to the muon flux from the four mesons using the pika model
}
\end{center}
\end{figure}

The charge ratio of muons from pion decay is $r_\pi={f_\pi}/(1-{f_\pi})$ 
and from kaon decay is ${r_K}={f_K}/(1-f_K)$.  We will refer to the
implications of Equation~\ref{eq:pika} with energy independent parameters
as the ``pika" model.
There are several interesting features of this model:
\begin{enumerate}
\item The relative intensity of cosmic ray pions and kaons that contribute to muon production 
can be extracted from surface and underground muon charge ratio experiments.
\item The muon charge ratio does not depend upon the muon energy and the zenith angle separately, 
but on the product $\ecos$. 
 This product of terms controls the relative portions of interaction
and decay for both pions and kaons.  At a fixed value of $\ecos$ , the intensity ratio 
of muons from pions and kaons is constant. 
\item There are contributions to the muon charge ratio $r_\mu$ from both pions and kaons from 1 GeV to 10 TeV. 
The contribution to the ratio from kaons does not vanish at GeV energies - it is just smaller
than at TeV energies. This effect can be seen in 
Figure~\ref{fig:mesoncontribution} which displays the
contribution to the muon flux from each of the four mesons.
\item
The pika model postulates an energy independent $r_\pi = {{\pi}^+}/{{\pi}^-}$ ratio 
related to
$f_{\pi}$, and energy independent $r_K = {K^+}/{K^-}$ ratio related to ${f_K}$, and an energy independent
 ${\pi}/K$ ratio embodied in the Gaisser constant 0.054.  
\item
Contributions from charm particle production have been ignored because their effect on the
ratio below 10 TeV is expected to be negligible.  
\end{enumerate}
\par
Where Feynman scaling is valid, the fraction $x=E_{meson}/E_{proton}$ for $\pi$ and
K secondaries does not depend upon $E_{proton}$.  Then $f_{\pi}$ and
 ${f_K}$ are also energy independent.
 Here we explore features of
our simple model, and in the next section we shall compare the qualitative features
of our model with full simulations of the charge ratio.
\par

Our parameterization yields additional insight into the sensitivity of various underground 
experiments which measure the
muon charge ratio.  Underground experiments can effectively observe a much smaller range
of $\ecos$ than a range of E.  To see this, consider a simple approximation
with five assumptions:
\begin{enumerate}
\item The parameter a in Equation ~\ref{eq:eloss} is constant
\item The parameter b in Equation ~\ref{eq:eloss} is constant and set to zero.
\item The earth/rock density above the underground detector is constant
\item The surface above the detector is flat
\item The MDM in the underground detector is negligible compared to the
energy loss from the surface to the detector.
\end{enumerate}
\par
The first four assumptions are equivalent to 
stating that the muon energy loss in the earth
is proportional to the overburden which only depends upon the muon zenith angle.  In that case
the energy loss through the overburden is 
${E_{loss}}={E_{min}/cos({\theta})}$, where $E_{min}$ is
 the minimum energy loss
for a vertical cosmic muon.  The fifth assumption implies that the underground energy
of the muons used for the charge ratio measurement is much smaller than their surface momentum.
\par
\begin{figure}[ht]
\begin{minipage}[t]{0.45\textwidth}

\begin{center}
\includegraphics*[width=6cm,angle=0,clip]{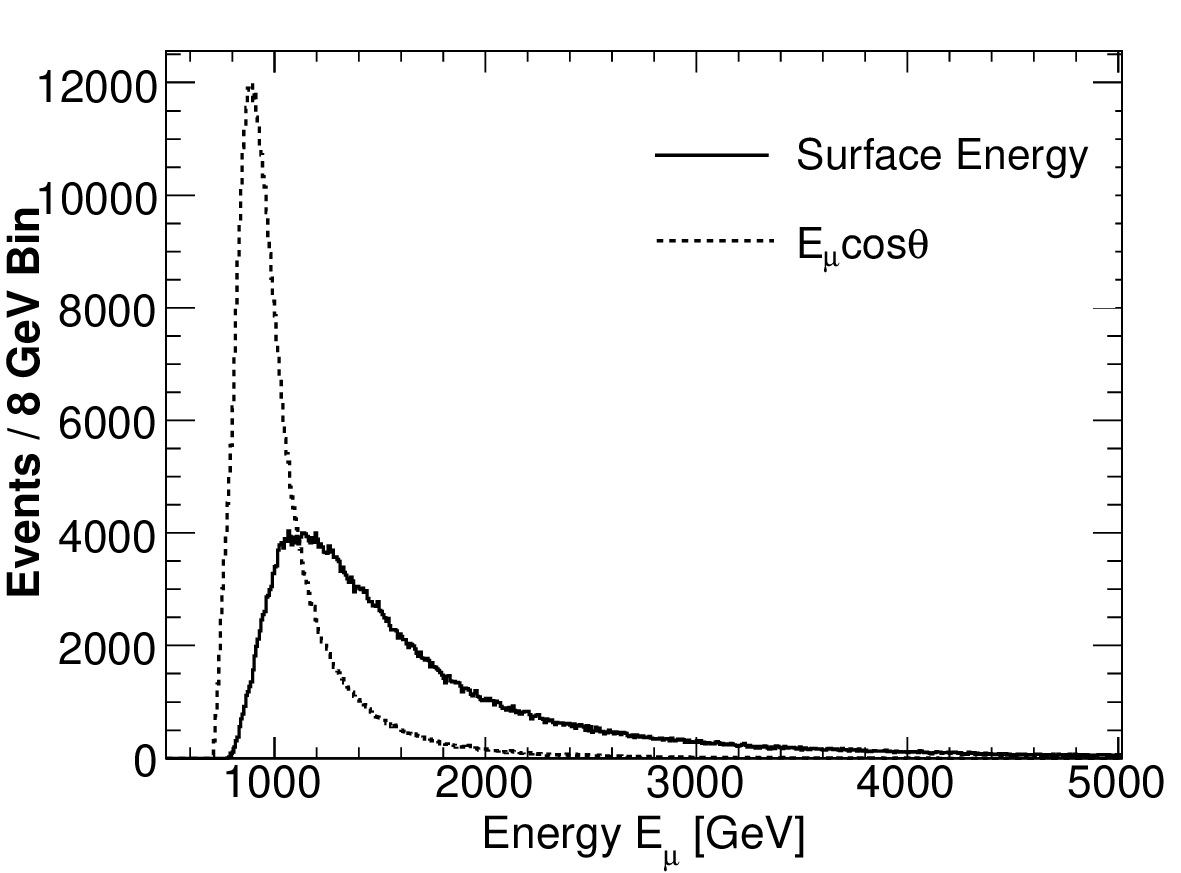}
\caption{\label {fig:farecos} 
Distribution of E and $\ec$ for MINOS data 
muons in the Far Detector, after cuts
(GeV).}
\end{center}
\end{minipage}
\hfill
\begin{minipage}[t]{0.45\textwidth}
\begin{center}
\includegraphics*[width=6cm,angle=0,clip]{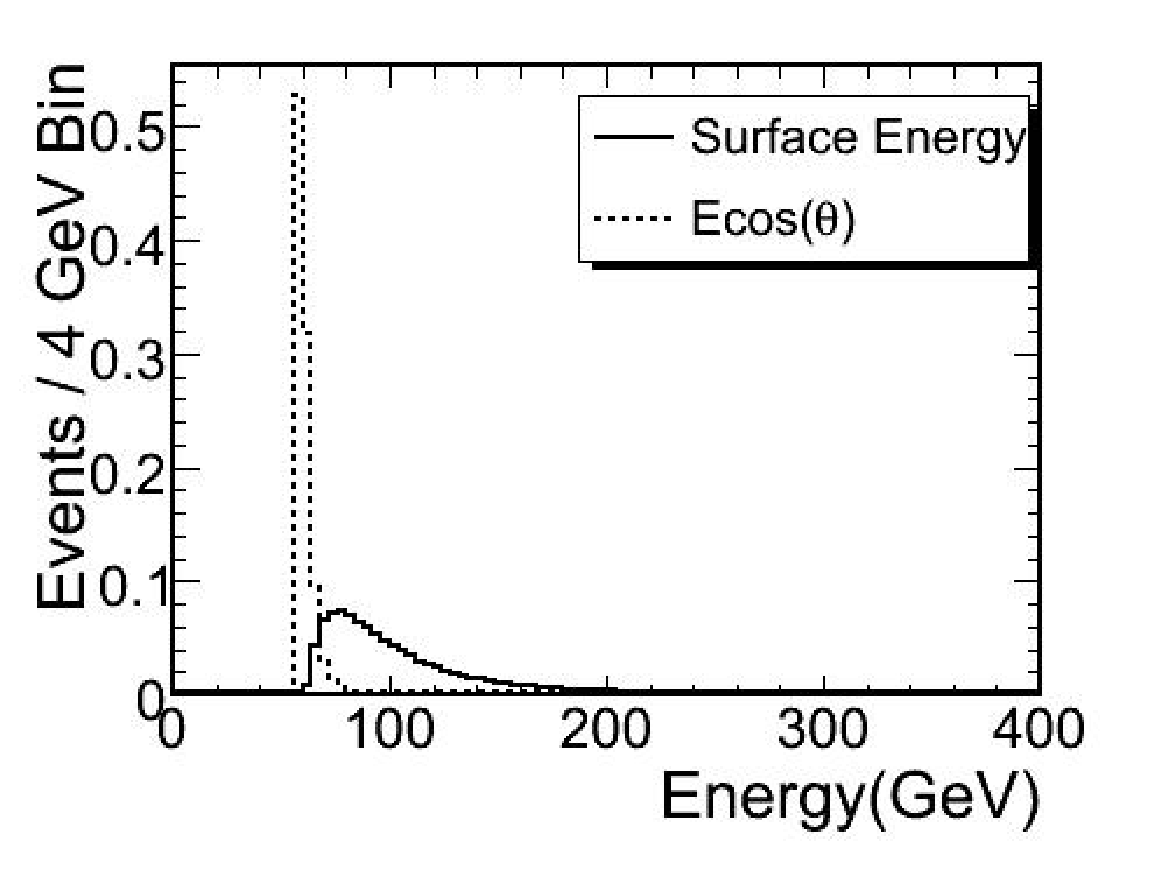}
\caption{\label {fig:nearecos} 
Distribution of E and $\ec$ for MINOS muons in the 
Near Detector, after cuts (GeV).}
\end{center}
\end{minipage}
\hfill
\end{figure}

With these five assumptions, the muon intensity distribution in $\ecos$ would be measured to be
a delta function at a value of $E_{min}$.  While at large zenith angles the surface energy
 increases due to the
1/$cos({\theta})$ dependence of the slant depth, the combined quantity $\ecos$ remains constant.
It is illuminating to compare this naive prediction 
to the actual distribution in the MINOS detectors when
no such assumptions are made.  These are
shown in Figures \ref{fig:farecos} and \ref{fig:nearecos}. 
Both of these MINOS 
$\ecos$ distributions are considerably narrower than the corresponding $E_{surface}$ distributions. 
The b(E) radiative term yields the largest contribution to the width
of the measured 
$\ecos$ distribution for the MINOS Far Detector. In the Near Detector distribution, 
the largest contribution is the larger ratio of maximum detectable momentum to energy loss
in the overburden.

\par
\begin{figure}
\begin{center}
\includegraphics [width=5.0in]{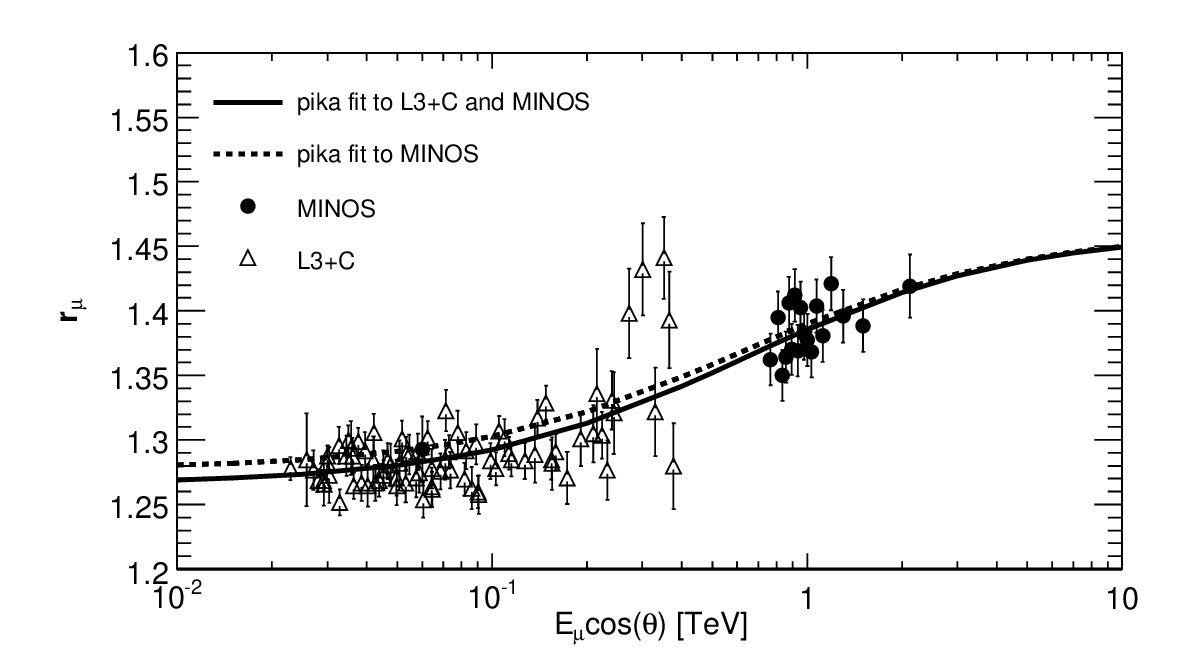}
\caption{\label{fig:pikafits-08}
 Pika model fitted to L3+C and MINOS (Near and Far) data sets
}
\end{center}
\end{figure}

We have used Equation~\ref{eq:pika}
and the measured muon charge ratio to study $r_\pi$ and $r_K$.
We have done chi-squared
fits in $\ecos$ to the MINOS Near Detector and Far 
Detector data, and to the L3+C data.  The
fit yields ${f_{\pi}}=0.5510 \pm 0.0006$ and
${f_K}=0.7006 \pm 0.0061$.  These values lead to 
a muon charge ratio from pion decay
of $r_\pi$ = 1.227 $\pm$ 0.003 and a muon charge ratio from charged kaon 
decay of $r_K$ = 2.34 $\pm$ 0.07.
The errors in the two parameters are highly correlated.  
This fit is shown in Figure \ref{fig:pikafits-08} along
with the MINOS (Near and Far) and L3+C data.  The agreement between 
the parametrization and the data is
excellent.   
\par
\begin{figure}
\begin{center}
\includegraphics [width=5.0in]{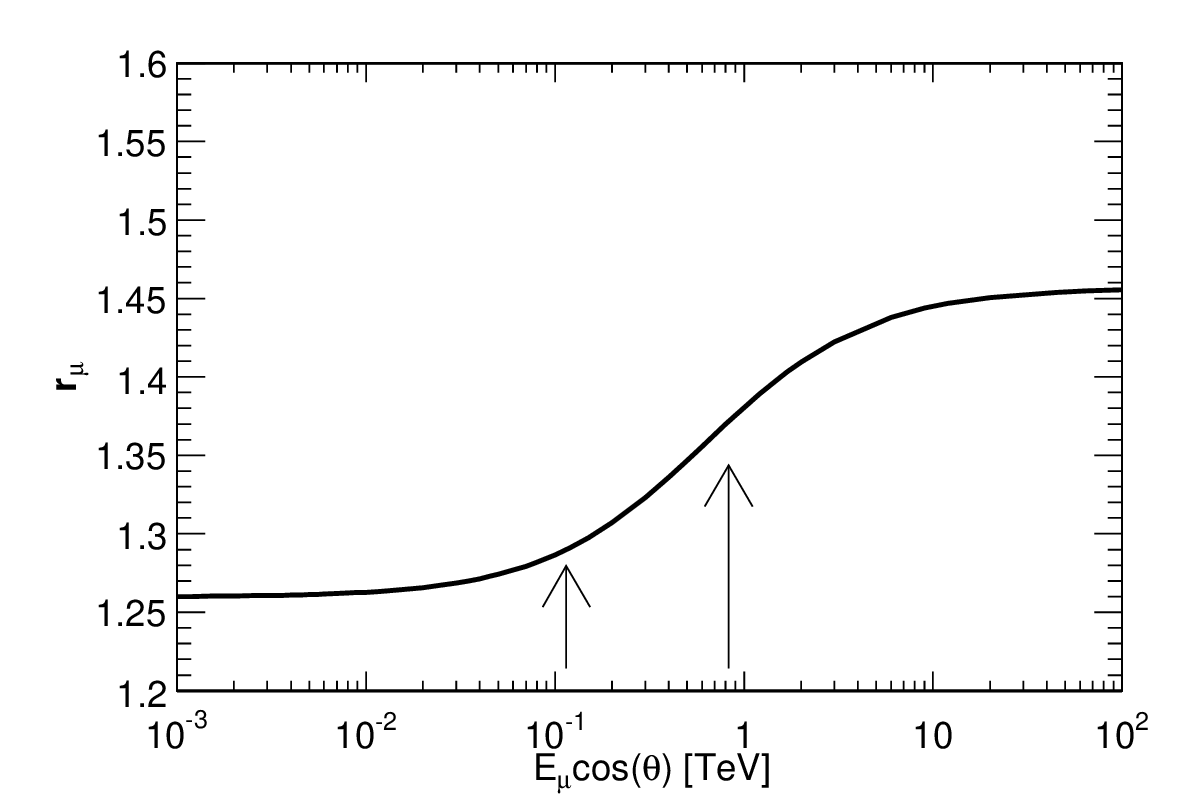}
\caption{\label{fig:pikalargeE}
The pika model's muon charge ratio for an extended range of energy.   The critical
energies for pion and kaon decay vs interaction are shown by the arrows.
}
\end{center}
\end{figure}

 Using the f values from this fit, we
plot Equation~\ref{eq:pika} for an extended range of muon energy
in Figure~\ref{fig:pikalargeE}.
The critical energies discussed above for pion and kaons are indicated by the arrows.
The low energy asymptotic ratio is 
$(f_\pi + \eta f_K)/[1-f_\pi +\eta(1-f_K)]$ = 1.26.
The high energy ratio is
$(\epsilon_\pi f_\pi + \epsilon_K \eta f_K)/
[\epsilon_\pi(1-f_\pi) + \epsilon_K \eta(1-f_K)]$ = 1.45.
The low and high energy asymptotic values both include
muons from $\pi$ decay and from K decay.

\par
The fit was repeated using just the smaller data sample of the 
MINOS Near Detector\cite{bib:near} and Far Detector.
 We obtain $f_{\pi}=0.5538 \pm 0.0070$ and ${f_K}=0.693 \pm 0.027$, 
which imply $r_{\pi}=1.241 \pm 0.035$
and ${r_K}=2.26 \pm 0.29$.  
\par
The parameterization we have used seems sufficient to represent the published 
data sets.  

\par
Our fits to $r_{\pi}$ gives values near expectations.  Our fit to ${r_K}={K^+}/{K^-}$ in atmospheric showers yield
values near 2.3.
It is clearly difficult to directly measure the atmospheric 
kaon charge ratio.  
Equation~\ref{eq:pika} provides a well defined parametrization for future studies of this subject.

\section{Comparison with Full Simulations and Calculations of Atmospheric Muons}
\label{sec:comparison}
\begin{figure}
\begin{center}
\includegraphics [width=5.0in]{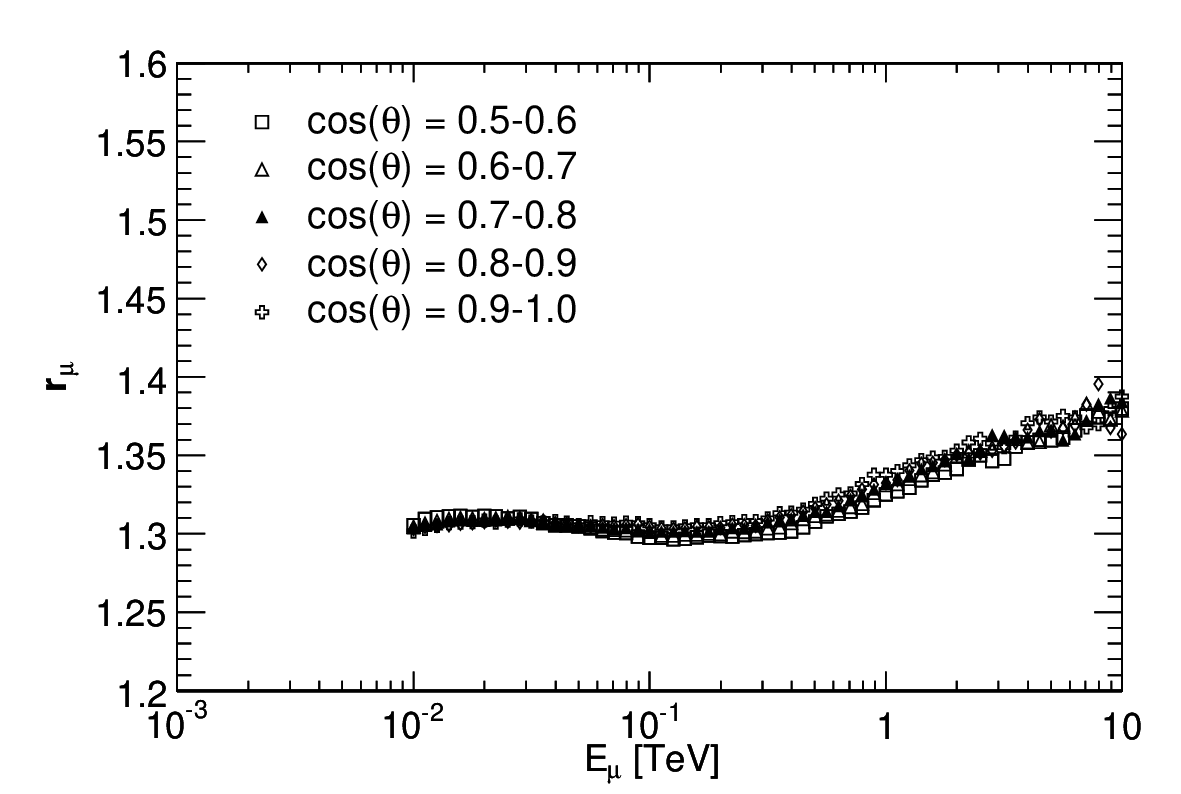}
\caption{\label{fig:hondaE}
Honda calculations of the muon charge ratio at 5 zenith angles
}
\end{center}
\end{figure}

\begin{figure}
\begin{center}
\includegraphics [width=5.0in]{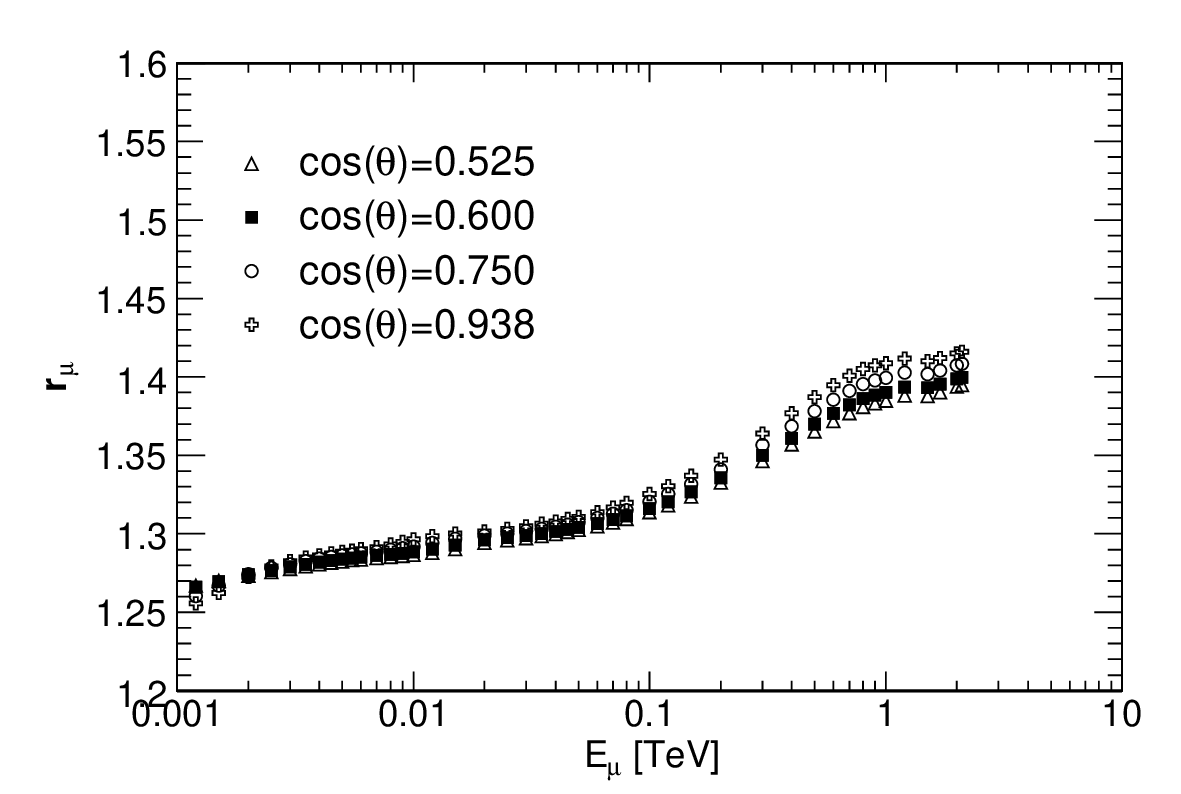}
\caption{\label{fig:cortE}
CORT calculations of the muon charge ratio at 4 zenith angles
}
\end{center}
\end{figure}

\begin{figure}
\begin{center}
\includegraphics [width=5.0in]{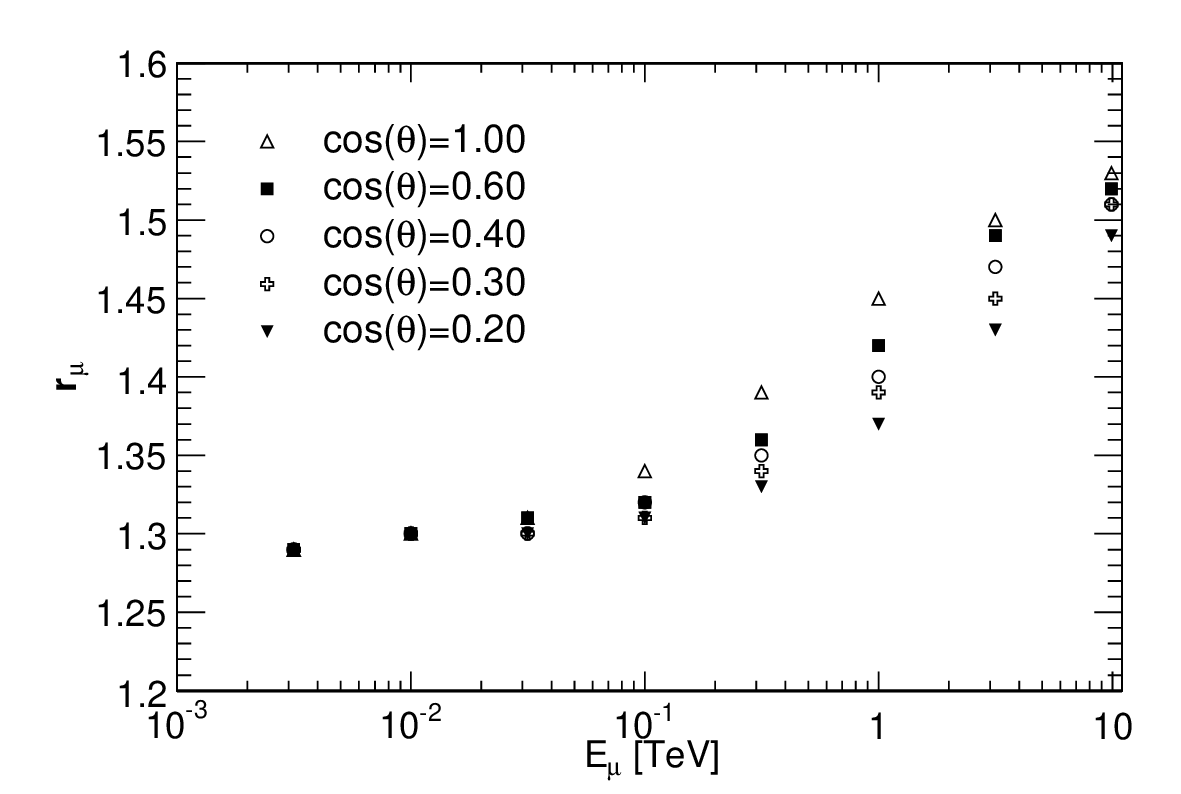}
\caption{\label{fig:lipariE}
Lipari calculations of the muon charge ratio at 5 zenith angles
}
\end{center}
\end{figure}

Full simulations of the muon charge ratio have been published by Honda \cite{bib:honda} and
CORT \cite{bib:naumov}; a full analytical calculation of the ratio has been published by Lipari\cite{bib:lipari}. 
These model results of the ratio versus the muon surface energy $\emu$ 
are displayed in Figures \ref{fig:hondaE}, \ref{fig:cortE}, and 
\ref{fig:lipariE} for various ranges of $\cos(\theta)$. The Honda results show small
dependencies upon the zenith angle, and an interesting dip in the ratio near 200 GeV.
The CORT results show a larger dependency upon the zenith angle as the energy
increases.  The Lipari results show even a larger dependency upon zenith angle
at energies above 10 GeV.
The results from the models differ noticeably from each other and from the MINOS
and L3+C data sets.

In our analysis, we do investigate (1)
the dependency of these charge ratio simulations/calculation on the
$\ecos$ variable discussed in the previous section of this paper,
and (2) their consistency with the pika model of the charge ratio.
Our observation will be that $\ec$  is a more useful variable for
ratio analysis than just the muon energy $\emu$.  

\par
\begin{figure}
\begin{center}
\includegraphics [width=5.0in]{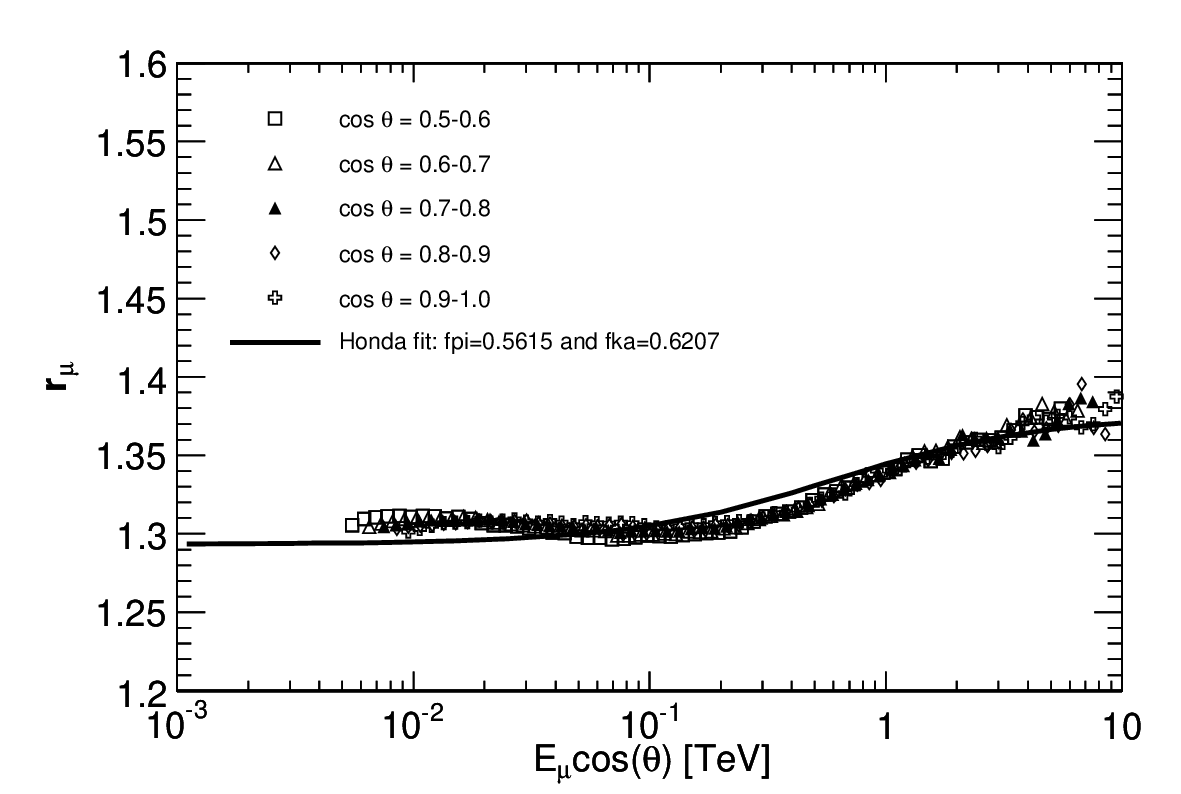}
\caption{\label{fig:hondaEcos}
Honda calculations of the muon charge ratio vs $\ecos$ at 5 zenith angles.
The curve is a fit of these simulation results to the pika model.
}
\end{center}
\end{figure}

  
Figure \ref{fig:hondaEcos} displays Honda's muon charge ratio simulations as a
function of  $\ec$ using published values from five 
ranges of $\cos({\theta})$ from 0.3 to 1.0.  
For a wide range of the variable $\ec$, the 
charge ratio is nearly independent of $\cos({\theta})$,
consistent with the representation of the pika model.

\begin{figure}
\begin{center}
\includegraphics [width=5.0in]{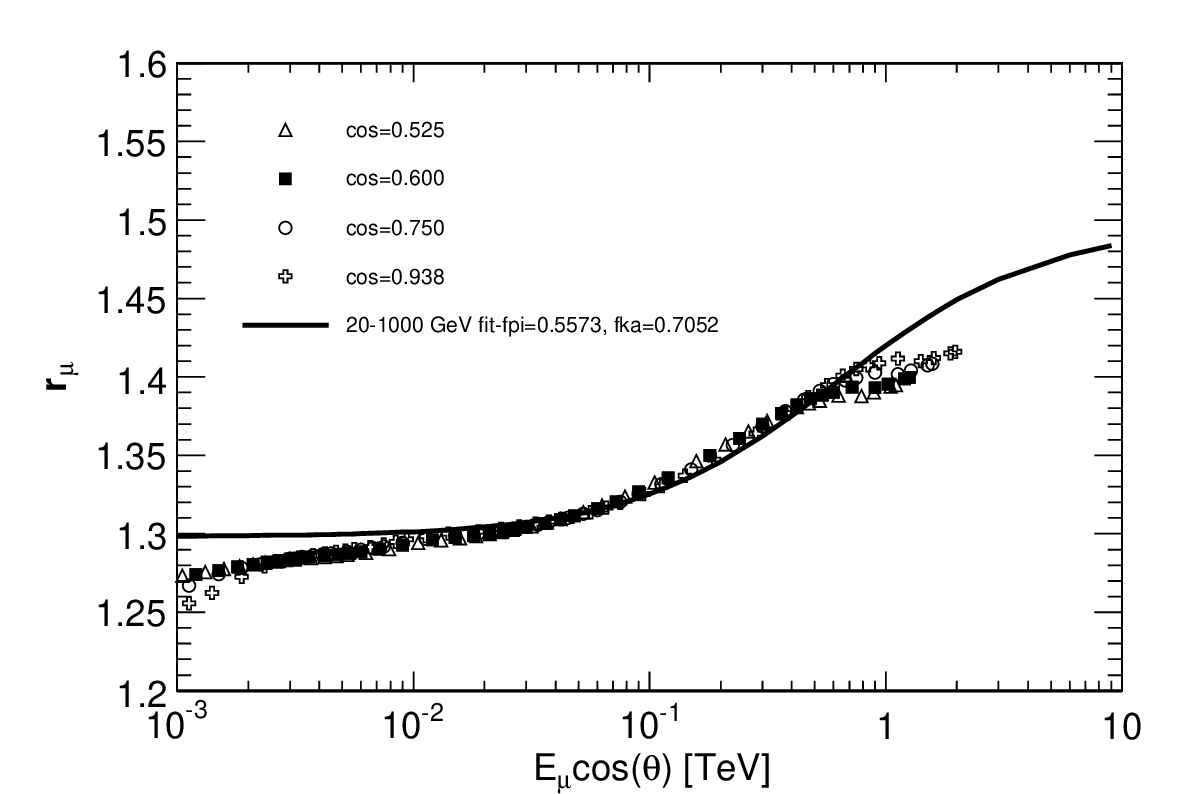}
\caption{\label{fig:cortEcos}
CORT calculations of the  muon charge ratio vs $\ecos$ at 4 zenith angles. 
The curve is a fit of these simulation results to the pika model.
}
\end{center}
\end{figure}

Figure \ref{fig:cortEcos} displays the corresponding CORT theory simulations of
the charge ratio as a function of $\ec$ using four published values of 
$\cos({\theta})$.  In the $\ecos$
range from 5 GeV to 600 GeV, 
the simulations show minimal dependency on $\cos( {\theta})$.  
Above 600 GeV, the charge ratio model results do increase with increasing values of $\cos({\theta})$. 
The prediction of the pika model that the ratio is independent of $\ec$ is consistent with the
CORT simulation for a remarkable range of $\ec$.

\begin{figure}
\begin{center}
\includegraphics [width=5.0in]{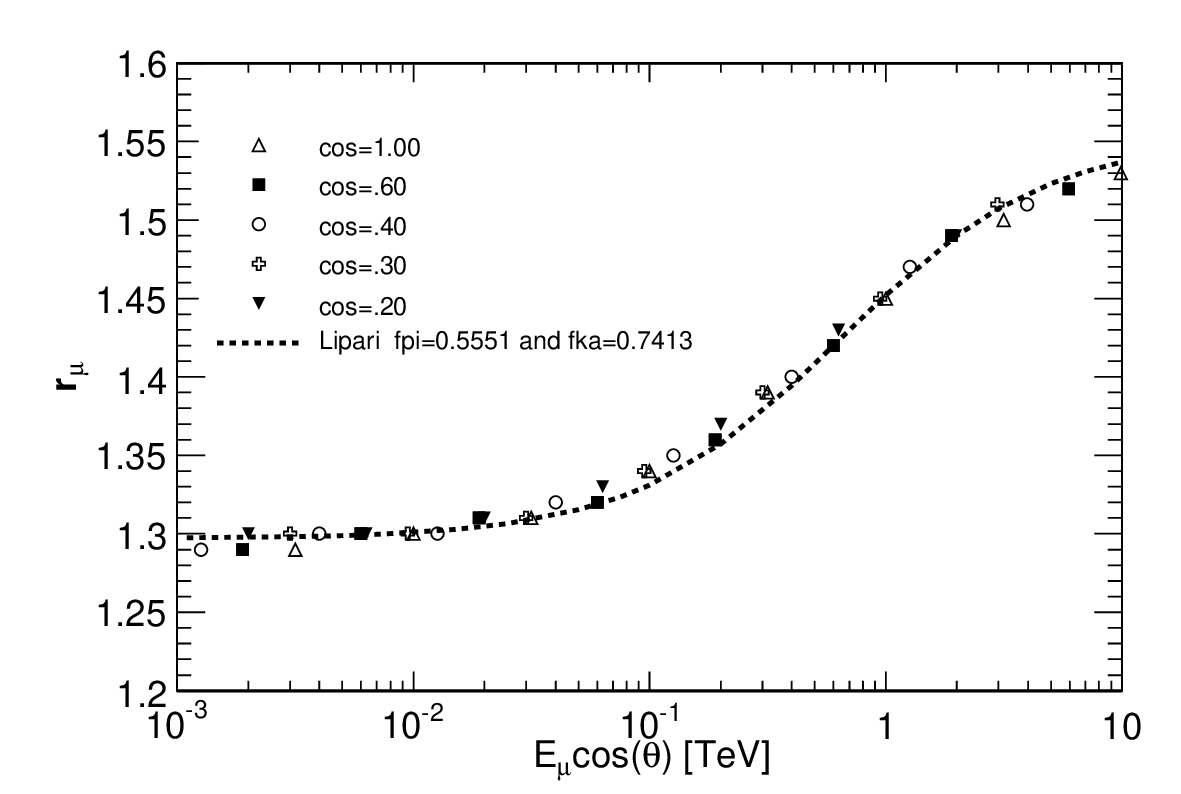}
\caption{\label{fig:lipariEcos}
Lipari calculations of the muon charge ratio vs $\ecos$ at 5 zenith angles.
The curve is a fit of this calculation's results to the pika model.
}
\end{center}
\end{figure}

Figure \ref{fig:lipariEcos} displays the Lipari calculation 
of the charge ratio as a 
function of $\ec$ for five values of $\cos({\theta})$ from 0.2 to 1.0. 
It is apparent that a significant portion of 
the ratio variation with $\cos({\theta})$ is minimal
 when the ratio is displayed as a function of $\ecos$. Again, the pika model prediction
 of a charge ratio independent of $\ec$ is consistent with the
Lipari calculation over four orders of magnitude in $\ec$. This agreement is to be
expected and is due to Lipari explicitly using the identical $\ec$ dependency
in his muon flux calculation, as Gaisser does in his model\cite{bib:gaisser} 
(our Equation~\ref{eq:gaisser}).
 \par
The observed dependence of the above three models on $\ecos$ provides
support for analyzing the charge ratio data using the pika formula. Of course
it would be interesting to
understand the source of the remaining variations in the simulations/calculation as a function of 
$\ecos$.
\par
Next we quantitatively compare the pika formula to the three simulations/calculation discussed above.
Two parameter chi-squared fits were performed
to the simulation's/calculation ratio values assuming equal uncertainties on each point within a model.
The best fit for each model is
shown as a smooth line in the corresponding figure.
For the CORT model, the pika model fits well the simulation from 20 GeV to
600 GeV at all angles.  The fit yields ${f_{\pi}}=0.557$ and ${f_K}=0.705$.
For the Honda simulation, the pika model can not reproduce the simulation's dip at $\sim$200 GeV,
although it does 
describe the simulation's overall energy dependency from 5 GeV to 5 TeV. The Honda fit yields
${f_{\pi}}=0.5615$ and ${f_K}=0.6207$.
For the Lipari calculation, the pika model must of course reproduce well the full calculation from 3 GeV to 10 TeV due to Lipari's usage of $\ec$. 
The fit yields ${f_{\pi}}=0.5551$ and ${f_K}=0.7413$. 
To summarize, we conclude there is physics content in 
using the variable $\ec$ in studying the
results of these three simulations/calculations. However, the pika formula can not reproduce the full richness 
of these simulations,
and of course the fitted pika parameters are significantly different among
the three simulations.
\par
\begin{figure}
\begin{center}
\includegraphics [width=5.0in]{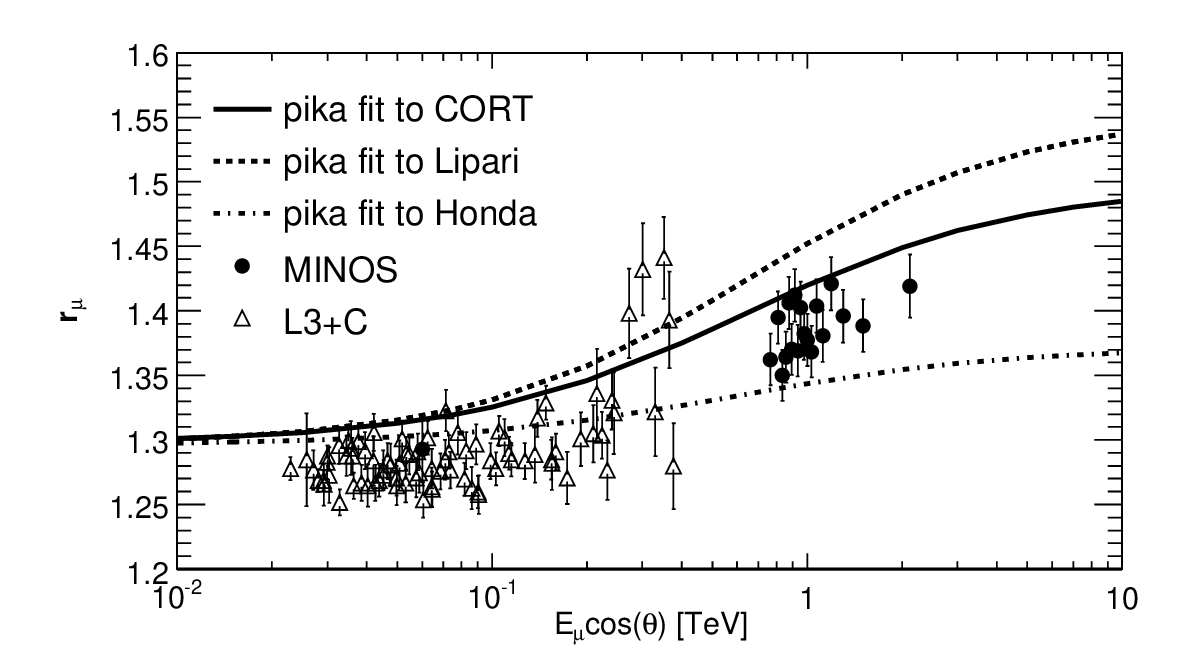}
\caption{\label{fig:3models}
 Fits of the pika formula to the three models of the charge ratio vs $\ecos$ compared 
 to L3+C and MINOS data sets.
}
\end{center}
\end{figure}

To complete this study, we have compared the L3+C and MINOS charge ratio 
data sets to the same three simulations/calculation.
So as to concisely represent the three simulations/calculation, we have used the fits of the three models to the pika formula
to compare with data.  As seen in 
Figure~\ref{fig:3models}.  
none of the models (as parameterized by the pika
formula with varying degrees of success) 
agree with the high $\ecos$ cosmic ray data.
Comparing the pika
model results shown in Figure~\ref{fig:pikalargeE} and Figure~\ref{fig:3models}, it is apparent that
none of the three simulations/calculation are using optimal values of ${f_{\pi}}$ and ${f_K}$. This can be seen in 
Table~\ref{tab:fits} and is discussed further 
in Section~\ref{sec:disc}. 

\section{Effect of Helium on the Charge Ratio}
Another process that could affect the energy dependence of $r_\mu$ would
be a different spectral index for the heavier cosmic ray primary flux than
for the proton flux.  This would introduce an energy dependence to the
incoming proton to neutron ratio.
It is not well established that the energy dependence of the heavy 
primary intensities is different
than that for Hydrogen in the 
10 TeV energy range. For this study, we will use Gaisser and Honda's \cite{bib:glasserhonda} 
parameterizations of the primary flux as a function
primary component k with energy $E_k$, given as \par
\begin{equation}
{\phi}({E_k})=K({E_k/GeV} + b \times exp[-c\sqrt{(E_k/GeV)}])^{-{\alpha}}
\end{equation}
The parameters in the above equation are given in the 
Table~\ref{tab:spec} below. Note that
this reference has a spectral index decreasing 
slightly with increasing primary mass.

\begin{table}[ht]
{
\begin{center}
\begin{tabular}{|c||c||c|c|c|}

\hline
{Parameter} & ${\alpha}$ & $K (m^{2}~ s~ sr~ GeV)^{-1}$ & $b$ & $c$ \\

\hline\hline
Hydrogen & 2.74$\pm$0.01 & 14900$\pm$600 &  2.15 & 0.21 \\
He (A=4) & 2.64$\pm$0.01 & 600$\pm$30 & 1.25 & 0.14 \\
CNO (A=14) & 2.60$\pm$0.07 & 33.2$\pm$5 & 0.97 & 0.01 \\
\hline

\end{tabular}
\end{center}
}
\caption{\label{tab:spec} Primary flux parameters used in the text.}
\end{table}

\begin{figure}
\begin{center}
\includegraphics [width=5.0in]{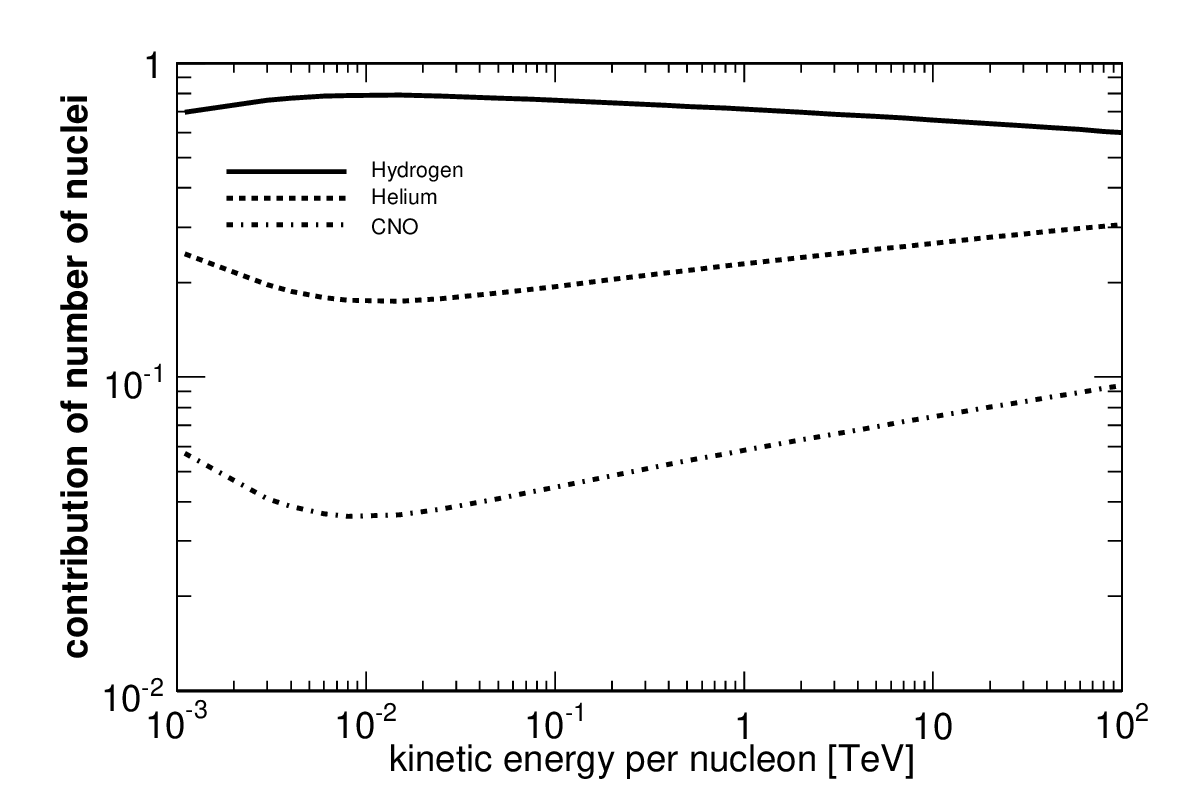}
\caption{\label{fig:hhecno}
Fractional contribution of nuclei to the all nucleon spectrum
}
\end{center}
\end{figure}

Figure~\ref{fig:hhecno} displays the contributions to
the flux as a function of the kinetic energy 
per nucleon.  Based on Monte Carlo calculations using CORSIKA, the 
mean surface energy 
of muons which reach 2100 mwe underground is 9\% of  the primary nucleon energy.
The mean fraction of the primary Helium energy, per nucleon, that is
transfered to the muon is 11\%.  
We will assume the fraction
of the energy transfer for primary carbon, nitrogen, and oxygen to be
identical to that of Helium.
In the following calculations, the energy profiles of the protons and Helium
from Monte Carlo are
 used instead of these average values, although the effect is not large.

\begin{figure}
\begin{center}
\includegraphics [width=5.0in]{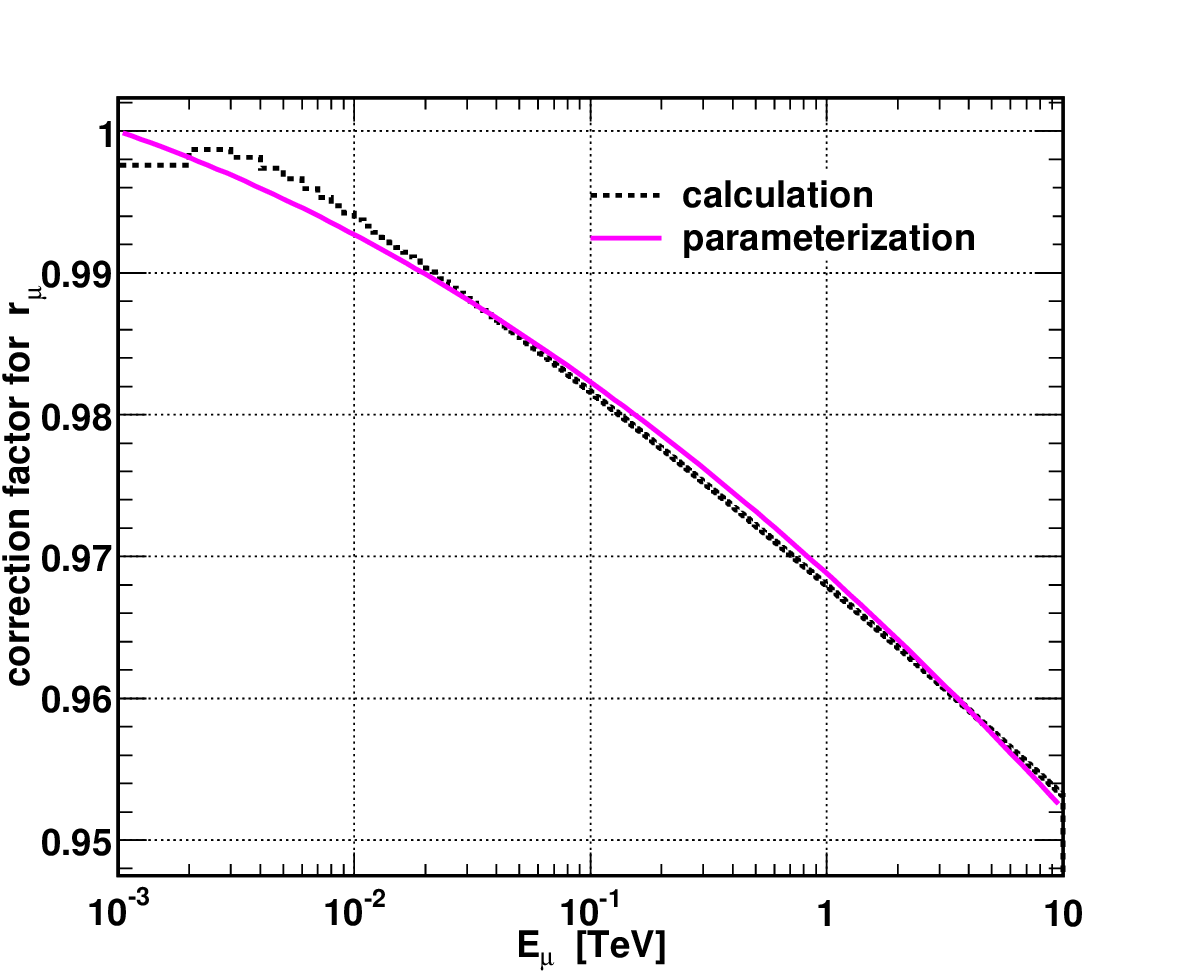}
\caption{\label{fig:fracchange}
Correction factor for 
$r_\mu$ assuming a different energy spectrum 
for Helium from pion production only.}
\end{center}
\end{figure}

We will calculate the corrections to the fitted $f_\pi$ and $f_K$
parameters in Equation~\ref{eq:pika}
to account for the possible energy dependency of the incoming proton
to neutron ratio (due to the Helium and CNO flux).  We use a model
to estimate the impact of both pion and kaon production 
by protons and neutrons.  We assume two contributions for meson
production:  pair production of equal number of positive and negative
mesons, and leading particle production using quark counting to estimate
the excess.  This calculation will use meson charge ratios near the
previous fit values.  We start with $r_\pi = 1.25$ and $r_K = 2.61$ and
see how much they change due to the different spectral index for
H and He. 

\par First we will model the change in the muon charge ratio due to just pion decay
when Helium and CNO are added to the primary proton flux.  
Using CORSIKA separately for cosmic ray protons and Helium of
energy $E_{CR}$, we have calculated
weights $w^i$ in 11 bins of the ratio $E_\mu/E_{CR}$.
In a simple quark model
the leading proton and neutron particle in the forward direction will give rise to
a charge ratio of 1.00 for symmetric nuclei A-Z=Z. 

Setting the muon charge ratio
equal to 1.25 at 1 GeV and fixing the 
normalization of the Hydrogen flux to be 1.317
(in order to get back the charge ratio of 1.25 at 1 GeV), one has for the
muon charge ratio

\begin{equation}
\begin{split}
r^{\pi}(E_{\mu}) =
 \sum_{i=0}^{10} [
{0.5 \times (w_p^i + w_{He}^i)}  \times  \\
\frac{ w_p^i  1.317  \times \phi_p(E_{CR}) + w_{He}^i  1.0 \times 4 
 \phi_{He}(E_{CR}) +
w_{He}^i  1.0 \times 14  \phi_{CNO}(E_{CR}) }
{ w_p^i  \phi_p(E_{CR}) + w_{He}^i  4  \phi_{He}(E_{CR}) + w_{He}^i  
14  \phi_{CNO}(E_{CR}) } ] \\
\end{split}
\end{equation}
The 1.0 values are due to the equal production rates of leading $\pi^+$
and $\pi^-$ in our quark counting model.
\par
A subtlety of our notation is that the $\pi$ (and K) superscript represents
the energy dependent correction to the charge ratio arising from
the different possible spectral index of heavy cosmic ray primaries,
while the subscript represents the assumed energy independent meson
ratio.
Figure~\ref{fig:fracchange} displays the corresponding fractional change in the charge ratio as a
function of the muon surface energy.  Note that the effect is about a
3 percent
reduction at 1 TeV.  This dependency can be fit to a polynomial in log(E),
also shown in Figure~\ref{fig:fracchange},
with the result

\begin{equation}
\label{eq:pi}
r^{\pi}(E_{\mu})/1.25 = 1-0.00575{\times} log_{10}(E_{\mu}/GeV) - 0.00155 {\times}
{{log^2}_{10}(E_{\mu}/GeV)}
\end{equation} 

Next we include  the contribution from kaon production by the primaries.  
The cross section for associated production of $\Lambda K^+$ is much
larger than that for $K^+K^-$ pairs for both incident protons and neutrons.
Again
assuming a quark model for the leading kaon in the forward direction from
incident protons and neutrons, one only gets leading $K^+$, but twice
as many from protons than from neutrons corresponding to their number
of $u$ quarks.
\begin{equation}
{p(uud) + air} {\rightarrow} K^{+}(u\bar{s}) + {\cdots}
\end{equation}
and
\begin{equation}
{n(ddu) + air }{\rightarrow} K^{+}(u\bar{s}) + {\cdots}
\end{equation}
In this model, the positive kaon excess originating from
protons ($r_K-1$) is twice as large as the excess originating from neutrons.
For symmetric nuclei with
A-Z=Z this leads to a charge ratio
\begin{equation}
\label{eq:19}
1 + \frac{[1+1/2]}{2} (r_K-1) = 1/4 + 3 r_K/4
\end{equation}
We further assume that the muon charge ratio at 1~TeV is 1.374 (near the
MINOS value).  This then yields
\begin{equation}
\begin{split}
r^{K}(E_{\mu}) =
 \sum_{i=0}^{10} [
{0.5 \times (w_p^i + w_{He}^i)}  \times  \\
\frac{ w_p^i  2.7  \times \phi_p(E_{CR}) + w_{He}^i  2.275 \times 4 
 \phi_{He}(E_{CR}) +
w_{He}^i  2.275 \times 14  \phi_{CNO}(E_{CR}) }
{ w_p^i  \phi_p(E_{CR}) + w_{He}^i  4  \phi_{He}(E_{CR}) + w_{He}^i  
14  \phi_{CNO}(E_{CR}) } ] \\
\end{split}
\end{equation}
The value 2.275 is due to Equation~\ref{eq:19}.  The leading particle
positive excess dominates over $K^+K^-$ pair production, unlike in the
pion case.  
Again we parameterize the dependency in $log_{10}{E_{\mu}}$ from Equation~\ref{eq:pi}
such that 
\begin{equation}
(1-r^K(E_{\mu})/2.61) = 2/3 \times (1-r^\pi(E_{\mu})/1.25)
\end{equation}
\par
The effect on the muon charge ratio due to the heavy ions requires that the pion and 
kaon fractions be modified as 
follows:
\begin{equation}
f^{*}_{\pi}(E_{\mu})=1/(1+1/r_{\pi} {\times} r^{\pi}(E_{\mu}))
\end{equation} 
\begin{equation}
f^{*}_K(E_{\mu})=1/(1+1/r_K {\times} r^{K}(E_{\mu})) 
\end{equation}
\par
\begin{figure}
\begin{center}
\includegraphics [width=5.0in]{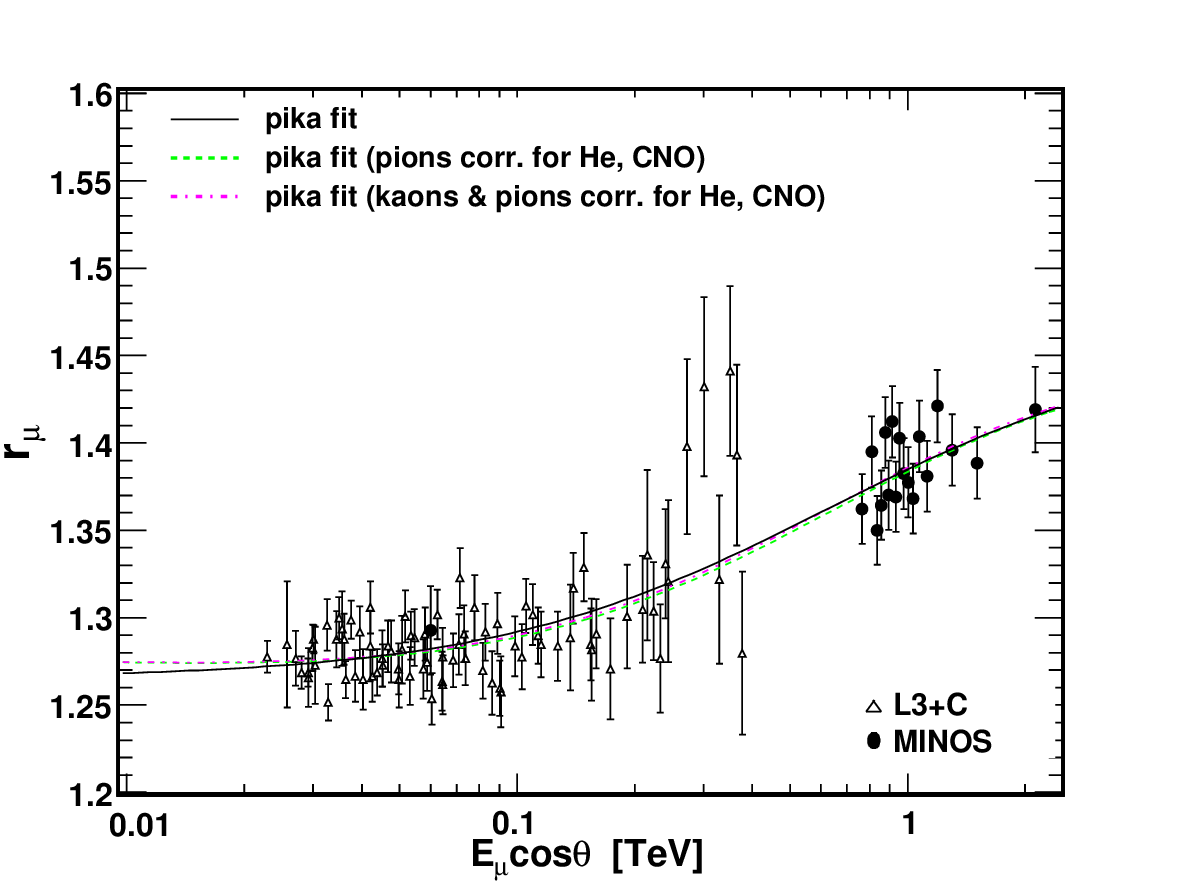}
\caption{\label{fig:Heliumratio}
Fit to the meson charge ratios modifying the pika model
with the energy dependence of Helium and CNO production.}
\end{center}
\end{figure}

Including these contributions from heavy primaries does have an impact 
on the calculation of parameters of the pika model.
The charge ratio no longer depends just upon $\ecc$.  While this effect is small
it is certainly present and can be accounted for in modeling.
Figure~\ref{fig:Heliumratio} shows the effect of the heavy primaries on the two parameters of the pika model.
The fit to the data is almost indistinguishable from the previous fit.  
Note that an increasing heavy primary fraction at high energy will decrease the
charge ratio, so to fit to the high MINOS points, a larger value of $r_K$
is required.
It is clear that simulations of the muon charge ratio need to include the 
possible different energy dependence of heavy primaries.

We repeat that the choice of spectral index in Table~\ref{tab:spec} is to
illustrate the size of a possible effect.  If it turns out that the
spectral index is independent of chemical composition, the effect
described in this section will not exist.

\section{Discussion}
\label{sec:disc}

We have considered the change in the charge ratio from the
fact that $\mu^+$ lose slightly more energy than $\mu^-$
while penetrating the overburden of
an underground detector, both for
ionization and catastrophic energy loss.  For catastrophic
energy loss, the effect is negligible, even for the small
errors in MINOS.  For ionization, we have calculated that
the 0.15\% difference in energy loss leads to a 0.6\% difference
in measured $r_\mu$ at 2000 mwe.  
This effect was mentioned
in Reference \cite{bib:prd} but not used to correct the
reported charge ratio there.  In this paper, we correct the MINOS
data to slightly higher values.   For the Far Detector data reported
in Reference~\cite{bib:prd}, this is a + 0.6\% correction, and for
the near detector at 400 mwe the correction is +0.37\%.
We estimate a +0.29\% correction for L3+C.

The central result of this paper is the development and
application of Equation~\ref{eq:pika} and the consequent relationship
of $r_\mu$ to $r_\pi$ and $r_K$.  We have used that equation
to study its consistency with several simulations/calculations, and to
make fits to data.  In Table~\ref{tab:fits}, we show the
fits for our parameters to the three more detailed models.  We also show
the fits to the data reported in References \cite{bib:prd}, \cite{bib:nd},
\cite{bib:l3} and \cite{bib:interpret}.
One can also compare these values to those used by 
Gaisser \cite{bib:gaisser} 
and Agrawal \cite{bib:agrawal};
assuming a proton primary cosmic ray flux (ignoring the Helium
component), these models use published pion and kaon production data from 
accelerator proton beams to specify the $r_\pi$ and $r_K$ parameters. In
particular, Gaisser specifies $r_\pi$ = 1.4 and $r_K$ =3.2; Agrawal specifies
$r_\pi$ = 1.35 and $r_K$ =2.92. 

\begin{table}[hbt] 
\centering 
 
{\begin{tabular}{|l|l||c|c|c|} \hline 
 & &  $r_K$ & $r_\pi$ \\ \hline
Calculation & Lipari &  2.87 & 1.25 \\ \hline
Simulation & CORT &  2.39 & 1.26 \\ \hline
Simulation & Honda &  1.63 & 1.28 \\ \hline
	Data fit & MINOS and L3  & 2.34 $\pm$ 0.07 & 1.227 $\pm$ 0.003 \\ \hline
Data fit & MINOS N+F  & 2.26 $\pm$ 0.29 & 1.241 $\pm$ 0.035 \\ \hline
Data fit & Helium $\gamma$ 2.74 $\rightarrow$ 2.64 &  2.73 $\pm$ 0.09 & 1.234
 $\pm$ 0.003  \\ \hline \hline

\hline 
\end{tabular}} 
\caption{\label{tab:fits}
Pika fits to $r_K$ and $r_\pi$ for a calculation, simulations, and data.}
\end{table}

\par  We have shown the importance of using $\ecos$ (instead of just energy) in 
the analysis of the atmospheric muon charge ratio for energies above 10 GeV. 
Understanding of the muon charge ratio would benefit if future experiments
 and theory simulations would provide results using $\ecos$.  
 The rise
in the charge ratio, which occurs between the $\pi$ and $K$ critical
energies of 115 GeV and 850 GeV, can be used to determine the meson charge
ratios $\kpm$ and $\pipm$.  This is an important new method
for obtaining information on these ratios. 

\par We further showed the effect on the charge ratio assuming that
there is a spectral
index for cosmic ray Helium nuclei which is different than the
spectral index for protons.  Using an index of 2.74 (2.64) for H (He),
we get a change in $r_K$ from 2.3 to 2.7 compared to an analysis where the
spectral indices were the same.

\par Both the data and the simulations seem to have 
consistent values for $r_\pi$.  
A general feature of our analysis is that the new MINOS
data by themselves suggest a lower value of $r_K$ than is
used in many of the simulations.  However, it is important to
note that the systematic effects considered
in this paper, which include remaining errors from randomization,
corrections for differences in dE/dx for $\mu^+$ and $\mu^-$ and
a possible lower spectral index for Helium, would all tend to raise
the fitted values of $r_K$.

\par Several effects which have not been explicitly considered
here are expected to be small.  These include the production of
muons from charm and other heavy particles, components of the
cosmic rays heavier than Helium and possible differences in their
spectra, and a variety of scaling violations which would have the
effect of making $f_\pi$ and $f_K$ energy dependent.  These 
effects need full simulations to evaluate fully; but it is 
important that the full simulations yield
the experimentally measured average values of these parameters.
In that context, the analysis presented
in this paper can be useful.

\section{Summary}

We have reviewed several factors which affect the 
measurement and interpretation of the muon charge ratio $r_\mu$
deep underground.  While many quantities in cosmic ray
physics are difficult to measure precisely, $r_\mu$ has been
measured in MINOS with a statistical accuracy better than
0.3\%.  In order to take advantage of this high precision, an
experiment must control systematic errors to a comparable
level, which is challenging.  We pointed out two kinds
of systematic error, those which might bias the measurement
of $r_\mu$, and those which randomize (and hence lower the
measured value from the true value).  Bias errors can be
canceled to high precision by using data with both
magnetic field polarities and Equation~\ref{eq:gmean}, 
and the success of this cancellation
can be checked for consistency.  Randomization errors, on
the other hand, require severe cuts that affect the statistical
precision.  Variables which may be related to randomization
effects can be identified and used to cut out data 
which may be affected by randomization.
However, there is no independent way to
determine if all such effects have been eliminated.  The
possibility exists that the true values of the charge ratio are
higher than those that have been reported.

The MDM is an important
parameter for an experiment with a magnet.  It affects both the calculation of the 
systematic error and the energy resolution, and also directly
limits the range of detected muon energies useful for measuring $r_\mu$
and $\ec$.

We note that to
fully explore
the rise in the charge ratio observed by MINOS, there is a need for
 additional precise ratio data in the $\ecos$ ranges of 0.2-0.8 TeV,
and above 3 TeV.  

\section{Acknowledgments}

This work was supported by the U.S. Department of Energy and Benedictine University.  
We would like to thank V. Naumov for introducing us to the issues involved in the predicted rise
of the charge ratio.  
We thank Tom Gaisser, Morihiro Honda, Paolo Lipari and Teresa Montaruli
for discussions on the ratio simulations.
We acknowledge the help and insight of Giles Barr, Thomas Fields, Jeff de Jong and Alec Habig.
We are grateful to Geoff Bodwin and Stanley Wojcicki for their contributions in understanding
issues involving energy loss.  
 We thank Eric Beall, Gavril Giurgiu, Eric Grashorn, Andrew Hoffman,
Sue Kasahara,
Stuart Mufson, Brian Rebel and Keith Ruddick for
 their many contributions.  And the support of the entire MINOS collaboration 
has been invaluable.

\end{document}